\documentclass[aps,pre,twocolumn,preprintnumbers,floatfix,nofootinbib]{revtex4-1}

\usepackage{amsmath}
\usepackage{amssymb}

\usepackage{graphicx}
\usepackage{subcaption}
\usepackage{epstopdf}
\usepackage{dcolumn}
\usepackage{bm}
\usepackage{marvosym}

\usepackage{color}
\definecolor{light-gray}{gray}{0.5}
\definecolor{blue}{rgb}{0.0,0.0,1.0}
\definecolor{green}{rgb}{0.0,0.5,0.0}
\definecolor{red}{rgb}{1.0,0.0,0.0}
\definecolor{cyan}{rgb}{0.0,0.75,0.75}
\definecolor{magenta}{rgb}{0.75,0.0,0.75}
\definecolor{yellow}{rgb}{0.75,0.75,0.0}

\newcommand{\avg}[1]{\langle{#1}\rangle}
\newcommand{\sdot}{\cdot}

\newcommand{\grad}{\bm \nabla}
\newcommand{\pd}{\partial}
\newcommand{\lrbig}[1]{\left( {#1} \right)}

\newcommand{\lt}{\left}
\newcommand{\rt}{\right}

\newcommand{\dd}{\mathrm{d}}

\begin{document}
\title{The origins of $k^{-2}$ spectrum in the decaying Taylor-Green magnetohydrodynamic turbulent flows}
\author{V. Dallas}
\email{vassilios.dallas@lps.ens.fr}
\author{A. Alexakis}
\affiliation{Laboratoire de Physique Statistique, Ecole Normale Superieure, 24 Rue Lhomond, 75231 Paris, France}

\begin{abstract}
We investigate the origins of $k^{-2}$ spectrum in a decaying Taylor-Green magnetohydrodynamic flow with zero large scale magnetic flux that was reported in \cite{leeetal10}. A possible candidate for this scaling exponent has been the weak turbulence phenomenology. 
From our numerical simulations, we observe that current sheets in the magnetic Taylor-Green flow are formed in regions of magnetic discontinuities. Based on this observation and by studying the influence of the current sheets on the energy spectrum, using a filtering technique, we argue that the discontinuities are responsible for the $-2$ power law scaling of the energy spectra of this flow.
\end{abstract}

\maketitle

\section{\label{sec:intro}Introduction}
In magnetohydrodynamic (MHD) turbulence there are several phenomenological theories \cite{biskamp03,zhouetal04,boldyrev06} competing as possible candidates for the interpretation of the power law exponent of the energy spectrum. Moreover, numerical simulations to date are unable to provide a definitive answer to this scaling. This has many implications; for example the energy dissipation rate, which is required to predict heating rates in solar and space physics \cite{marinoetal08}, is connected to the slope of the energy spectrum.

In freely decaying isotropic MHD turbulence, some simulations obtained $k^{-3/2}$ while others $k^{-5/3}$ scaling for the energy spectra \cite{mullergrappin05,mininnipouquet07}. Observations from astrophysical plasmas have shown that this difference in the power law scaling also exists for the measured energy spectra of the solar wind \cite{podestaetal07}. Recently, large resolution simulations by Lee et al. \cite{leeetal10} demonstrated $k^{-2}$, $k^{-5/3}$ and $k^{-3/2}$ total energy spectrum scalings for different initial conditions of the magnetic field. Hence, they showed dependence of the energy spectrum at the peak of dissipation on the initial conditions and consequently they suggested lack of universality in decaying MHD turbulent flows. The difference between $-5/3$ and $-3/2$ power laws is subtle enough (10\% difference) that an inertial range of an order of magnitude is not enough to make a clear distinction between them. However, a $-2$ scaling exponent can be more transparent at least in such 
high enough Reynolds numbers. Indications of $k^{-2}$ scaling are also reported for the magnetic energy spectrum measured in the magnetosphere of Jupiter \cite{sauretal02}.

At the time Lee et al \cite{leeetal10} interpreted the $k^{-2}$ spectrum in terms of weak turbulence (WT) theory that predicts this exponent for weakly interacting waves in the presence of strong uniform magnetic field. Here, we would like to emphasise that the WT scaling is for an anisotropic energy spectrum $E(k_\parallel,k_\perp) \propto f(k_\parallel) k_\perp^{-2}$ \cite{ngbhattacharjee97,galtieretal00}, where the indices $\parallel$ and $\perp$ indicate the direction parallel and perpendicular to an imposed large scale mean magnetic field $B_0$, respectively. 
In the simulations of \cite{leeetal10} no large scale magnetic field was applied but large scale magnetic
structures were formed that were assumed to play the role of $B_0$ locally.
In this paper, we investigate the origins of the $k^{-2}$ spectrum through direct numerical simulations (DNS) by reconsidering the insulating magnetic Taylor-Green (TG) initial condition used in \cite{leeetal10}.

The paper is organised as follows. All the necessary details on our DNS of decaying MHD turbulent flows are provided in section \ref{sec:dns}. Section \ref{sec:scalings} focuses at the scaling of the energy spectra of our flows and provides an outline of the WT phenomenology. In particular, we focus on the justification of the $k^{-2}$ scaling for the spectrum of the magnetic energy $E_b$. Based on clear indications from our DNS, that regions of high shear with abrupt changes in the direction of the magnetic field occur in the flow, we show that the $-2$ power law can be derived analytically without WT assumptions (see section \ref{sec:disc}). To further support our argument, in section \ref{sec:filter} we employ a filtering technique to assess if the $k^{-2}$ scaling originates from these strong shearing regions that manifest as discontinuities in the magnetic field or not. Finally, in section \ref{sec:end} we conclude by summarising our findings.

\section{\label{sec:dns}Numerical simulations}
\subsection{Governing equations \& numerical method}
In this study, we deal with the three-dimensional, incompressible MHD equations of fluid velocity $\bm u$ and magnetic field $\bm b$
\begin{align}
 & \pd_t \bm u = (\bm u \times \bm \omega) - \grad P + \nu \bm\Delta \bm u + (\bm j \times \bm b)
 \label{eq:ns} \\
 & \pd_t \bm b = \grad \times (\bm u \times \bm b) + \kappa \bm\Delta \bm b
 \label{eq:induction} \\
 & \grad \sdot \bm u = \grad \sdot \bm b = 0
 \label{eq:incomp}
\end{align}
with $\nu$ the kinematic viscosity, $\kappa$ the magnetic diffusivity,  $\bm \omega \equiv \grad \times \bm u$ the vorticity, $\bm j \equiv \grad \times \bm b$ the current density of the magnetic field and $P = p/\rho + \tfrac{1}{2}\bm u^2$ the fluid pressure, composed by the ratio of plasma pressure $p$ with the constant mass density $\rho$ and the hydrodynamic pressure $\tfrac{1}{2}\bm u^2$. The magnetic induction can be defined through a magnetic potential $\bm a$ as $\bm b \equiv \grad \times \bm a$ with $\grad \sdot \bm a = 0$. Note that magnetic induction has units of Alfv\'en velocity, i.e. $\bm b/\sqrt{\rho \mu_0}$, where $\mu_0 = (\kappa \sigma)^{-1}$ is the permeability of free space with $\sigma$ the electrical conductivity. For $\nu = \kappa = 0$, the total energy $E_t \equiv \frac{1}{2}\avg{|\bm u|^2 + |\bm b|^2} = E_u + E_b$, the magnetic helicity $H_b \equiv \avg{\bm u \sdot \bm b}$ and the cross helicity $H_c \equiv \avg{\bm a \sdot \bm b}$ are conserved in time, where the angle brackets 
$\avg{.}$ in this study denote spatial averages.

To numerically solve Eqs. \eqref{eq:ns}-\eqref{eq:incomp} we employ the standard pseudo-spectral method \cite{gottlieborszag77}, where each component of $\bm u$ and $\bm b$ is represented as truncated Galerkin expansions in terms of the Fourier basis. The non-linear terms are initially computed in physical space and then transformed to spectral space using fast-Fourier transforms \cite{fftw98}. Aliasing errors are removed using the 2/3 dealiasing rule, i.e. the maximum wavenumber is $k_{max} = N/3$, where $N$ is the number of grid points in each Cartesian coordinate of our periodic box with period $2\pi$. The non-linear terms along with the pressure term are computed in such a way that $\bm u$ and $\bm b$ are projected on to a divergence-free space so that Eqs. \eqref{eq:incomp} are satisfied \cite{mpicode05a}. The temporal integration of Eqs. \eqref{eq:ns} and \eqref{eq:induction} is performed using a second-order Runge-Kutta method. The code is parallelised using a hybrid parallelisation (MPI-OpenMP) 
scheme \
cite{hybridcode11}.

\subsection{\label{sec:IC}Initial conditions \& numerical parameters}
Based on the results of \cite{leeetal10}, we choose
the initial velocity field to be the Taylor-Green vortex \cite{taylorgreen37} defined as 
\begin{equation}
 \label{eq:TGu}
 \bm u_{TG}(\bm x) = u_0 (\sin x \cos y \cos z, -\cos x \sin y \cos z, 0)
\end{equation}
and the initial magnetic field to be a modification of the TG vortex, i.e. $\bm b_{TG} = -(b_0/u_0) \grad \times \bm u_{TG}$, which takes the following form
\begin{equation}
 \label{eq:TGbi}
 \bm b_{TG}(\bm x) = b_0 %(\cos x \sin y \sin z, \sin x \cos y \sin z, -2 \sin x \sin y \cos z)
 \begin{pmatrix}
  \cos x \sin y \sin z \\
  \sin x \cos y \sin z \\
  -2 \sin x \sin y \cos z
 \end{pmatrix}^T.
\end{equation}
The current density $\bm j_I$ is everywhere parallel to the faces of the sub-boxes $[0,\pi]^3$, called the impermeable boxes \cite{brachetetal83}, and thus considered as electrical insulators. Note that the magnetic and cross helicity are globally restricted due to the TG symmetries to vanish for all times.

This magnetic TG flow exhibits several intrinsic symmetries within the periodic box of size $[0,2\pi]^3$ (see also \cite{brachetetal83}). These are mirror (anti)symmetries about the planes $x=0$, $x=\pi$, $y=0$, $y=\pi$, $z=0$ and $z=\pi$ as well as rotational (anti)symmetries of angle $N\pi$ about the axes $(x,y,z)=(\tfrac{\pi}{2},y,\tfrac{\pi}{2})$ and $(x,\tfrac{\pi}{2},\tfrac{\pi}{2})$ and of angle $N\pi/2$ about the axis $(\tfrac{\pi}{2},\tfrac{\pi}{2},z)$ for $N \in \mathbb{Z}$. The above mentioned planes that possess mirror symmetries form the insulating faces of the impermeable boxes.

Note that Lee et al. \cite{leeetal10} enforced numerically these symmetries in order to gain substantial savings in both computing time and memory usage at a given Reynolds number. In contrast, our DNS of the magnetic TG flow was performed without imposing any symmetry constrains, allowing thus the turbulence to evolve freely. As it was observed in \cite{da13a}, where no symmetries where also imposed for the MHD TG flows, even for their highest Taylor Reynolds number simulations ($\sim \mathcal O(10^2)$), the TG vortex symmetries did not break within the time interval of reaching the peak of dissipation. This indicates that these symmetries are a strong property of the MHD equations, preserved by time evolution of the solutions (see also \cite{leeetal08}).

For comparison to the MHD TG flow, which carries special global restrictions due to the TG symmetries, we further consider a run with random initial conditions (run ``R'' hereafter). In order to obtain the broadest inertial range, run R is initialised by exciting wavenumbers with $|k| = 1$ and 2 with random phases. At $t = 0$, we ensure $H_b = H_c = 0$ as well as kinetic helicity $H_u \equiv \avg{\bm u \sdot \bm \omega} = 0$. During the time evolution magnetic and cross helicity remain zero for all times relative to the total energy. However, the kinetic helicity reaches an approximate value of $H_u\ell/E_t < 0.2$ at its absolute maximum over time but when dissipation is maximum $H_u\ell/E_t < 0.04$ and hence negligible.  

In the simulations we used unit magnetic Prandtl number ($\nu = \kappa$), and a grid of $N = 1024^3$ points. At time $t=0$ the fields are normalised such that $E_u = E_b = 0.125$, i.e. the kinetic and magnetic energies are in equipartition. All the necessary numerical parameters of our DNS are provided in Table \ref{tbl:dnsparam}.

\begin{table}[!ht]
  \caption{Numerical parameters of the DNS. The values presented are taken at the peak of total dissipation. Note that $k_{max} = N/3$, using the $2/3$ dealiasing rule.}
  \label{tbl:dnsparam}
%    \begin{ruledtabular}
\resizebox{0.5\textwidth}{!}
{
    \begin{tabular}{*{7}{@{}c}cc@{}c@{}}
%     \begin{tabular}{*{10}{c}}
     \hline
     \hline   
      \textbf{Run} & \textbf{N} & $\bm{\nu}$ & $\bm{Re_{\lambda_t}}$ & $\bm{L_t}$ & $\bm{\lambda_t}$ & $\bm{\eta_t}$ & $\bm{u'}$ & $\bm{b'}$ & $\bm {k_{max}\eta_t}$ \\
       & & $(\times 10^{-4})$ & & $(\times 10^{-1})$ & $(\times 10^{-1})$ & $(\times 10^{-3})$ & & & \\
     \hline
      TG & 1024 & 4.5 & 121.8 & 6.84 & 2.03 & 6.54 & 0.27 & 0.62 & 2.23 \\
      R  & 1024 & 4.0 & 164.7 & 7.95 & 1.83 & 6.18 & 0.36 & 0.49 & 2.11 \\
     \hline
     \hline 
    \end{tabular}
}
%   \end{ruledtabular}
\end{table}

The total, kinetic and magnetic integral length scales are defined respectively as
\begin{equation}
 L_{t,u,b} \equiv \frac{3\pi}{4}\frac{\int k^{-1}E_{t,u,b}(k)dk}{\int E_{t,u,b}(k)dk}
\end{equation}
and similarly applies for the Taylor scales
\begin{equation}
 \lambda_{t,u,b} \equiv \lrbig{5\frac{\int E_{t,u,b}(k)dk}{\int k^2E_{t,u,b}(k)dk}}^{1/2}.
\end{equation}
In Table \ref{tbl:dnsparam}, we report the $L_t$ and $\lambda_t$ as well as the Reynolds number based on the total Taylor length scale given by $Re_{\lambda_t} \equiv u' \lambda_t / \nu$. 
The rms velocity $u'$ is defined as
\begin{equation}
  u' \equiv \lt( \frac{2}{3}\int E_u(k)dk \rt)^{1/2}
\end{equation}
and similarly the rms magnetic field $b'$.
Finally, the smallest length scale in our flows is defined based on K41 scaling $\eta_t \equiv (\nu^3 / \epsilon_t)^{1/4}$, where $\epsilon_t = \nu \avg{|\bm \omega|^2} + \kappa \avg{|\bm j|^2}$ is the total dissipation. The time we address in our analysis is the moment of maximum dissipation, when the highest scale separation occurs $\eta_t \ll \ell \ll L_t$, where $\ell$ is a typical length scale in the inertial range. Thus, the values provided in Table \ref{tbl:dnsparam} correspond to that moment.

\section{\label{sec:scalings}Scaling of the energy spectra}
\begin{figure}[!ht]
  \begin{subfigure}{6cm}
   \includegraphics[width=\textwidth]{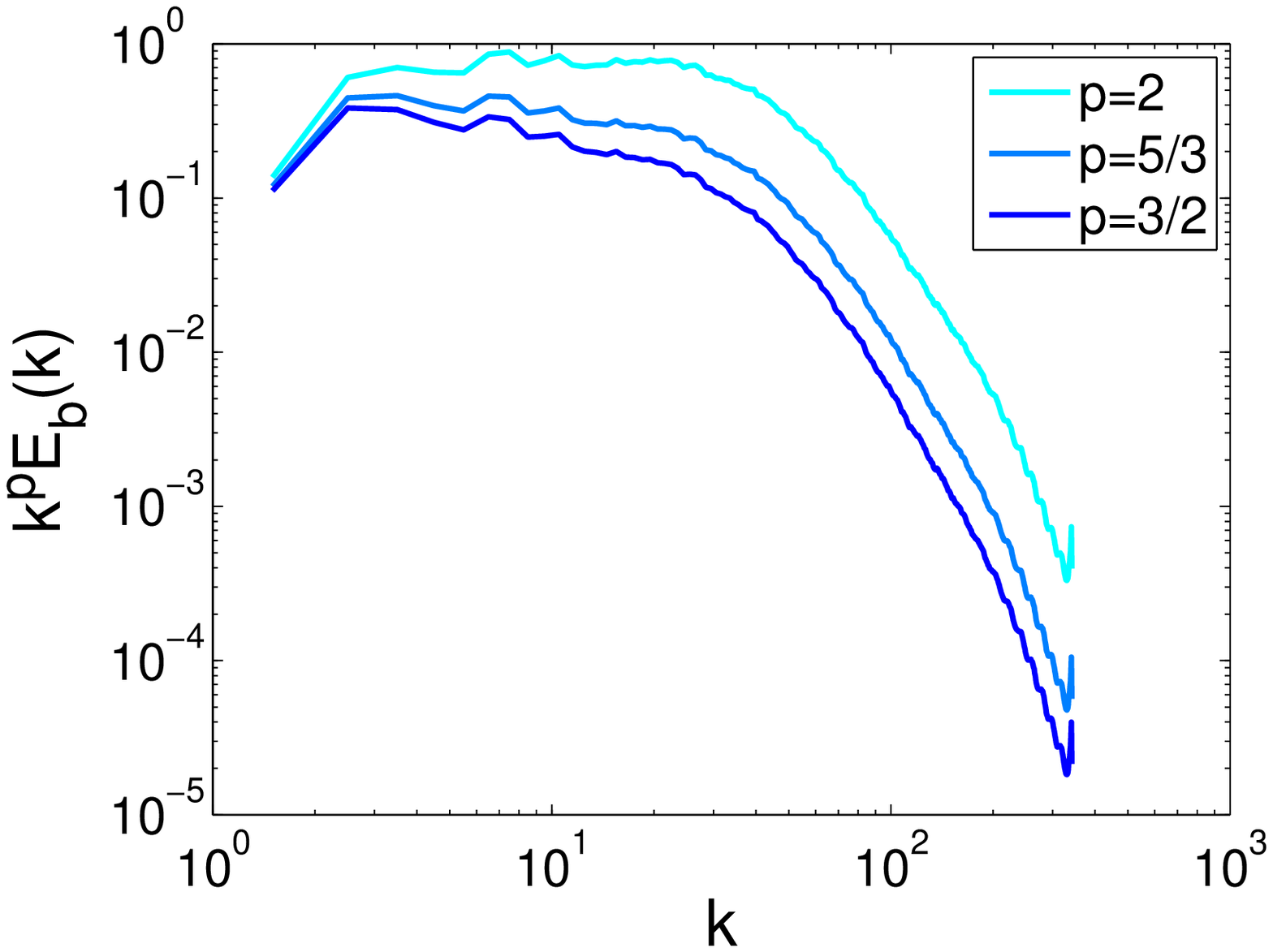} 
   \caption{}
  \end{subfigure}
  \begin{subfigure}{6cm}
   \includegraphics[width=\textwidth]{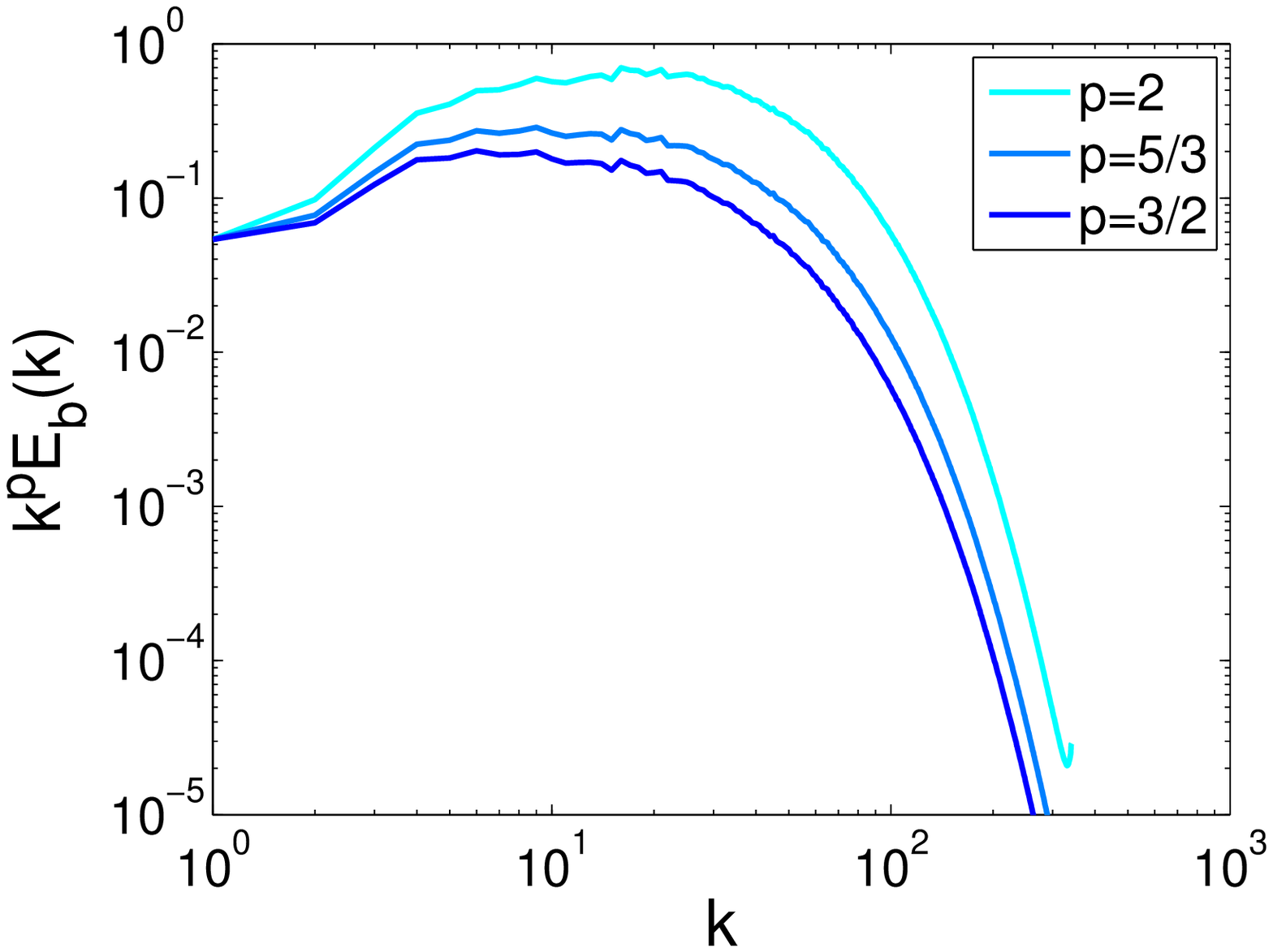}
   \caption{}
  \end{subfigure}
  \caption{(Color online) Three-dimensional compensated magnetic energy spectra $k^pE_b(k)$ with scaling exponents $p = 2,\; 5/3,\; 3/2$ for (a) run TG and (b) run R of Table \ref{tbl:dnsparam}.}
  \label{fig:Eb_spectra}
 \end{figure}

\begin{figure}[!ht]
  \begin{subfigure}{6cm}
   \includegraphics[width=\textwidth]{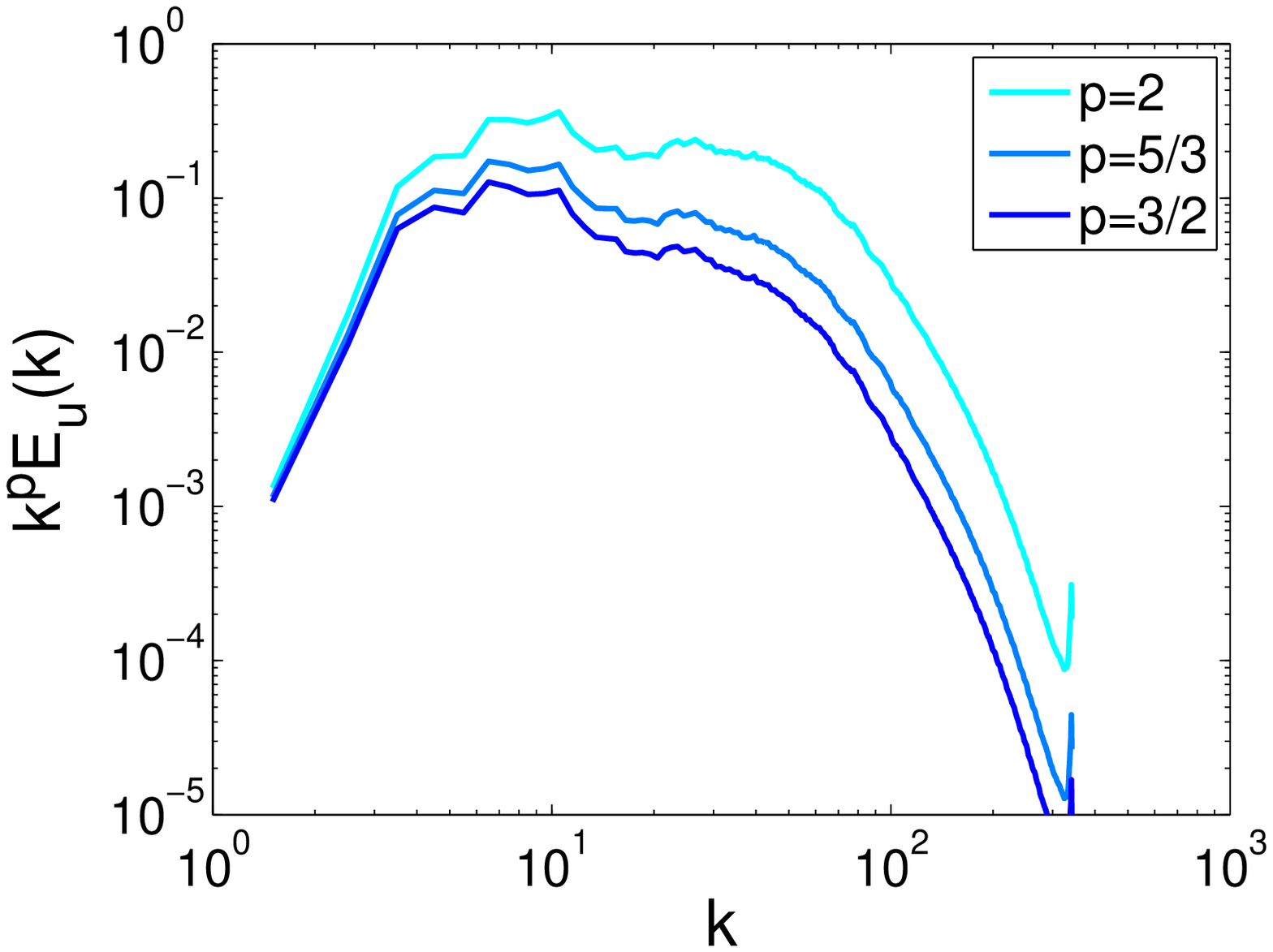} 
   \caption{}
  \end{subfigure}
  \begin{subfigure}{6cm}
   \includegraphics[width=\textwidth]{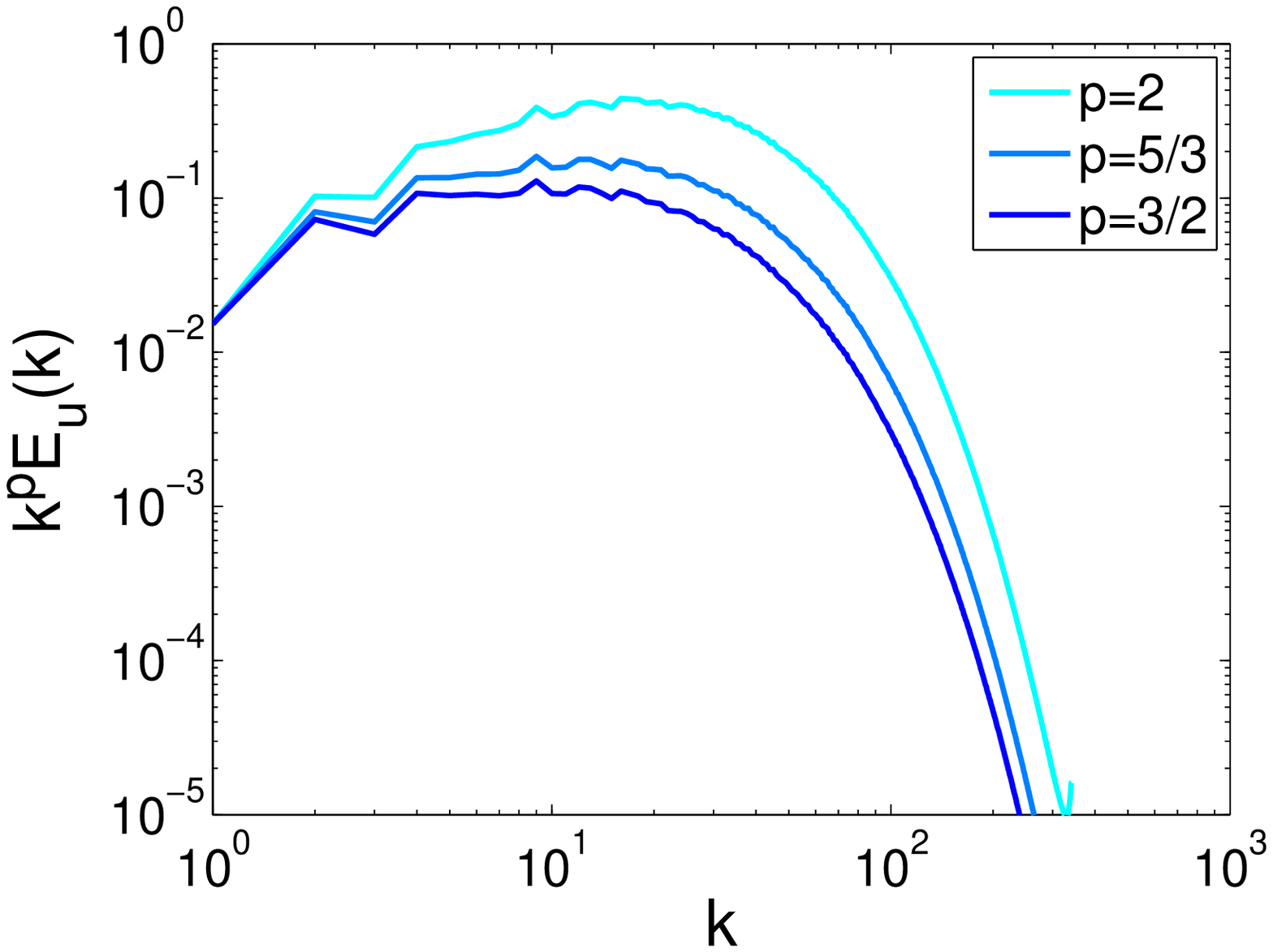}
   \caption{}
  \end{subfigure}
  \caption{(Color online) Three-dimensional compensated kinetic energy spectra $k^pE_u(k)$ with scaling exponents $p = 2,\; 5/3,\; 3/2$ for (a) run TG and (b) run R of Table \ref{tbl:dnsparam}.}
  \label{fig:Eu_spectra}
 \end{figure}

Figures \ref{fig:Eb_spectra} and \ref{fig:Eu_spectra} present the three-dimensional compensated magnetic and kinetic energy spectra, respectively, that we obtain for runs TG and R at the peak of dissipation. The spectra are compensated with the exponents $p = 2,\; 5/3$ and $3/2$. These are in summary the power law scaling exponents obtained in the various MHD turbulence phenomenologies based on weak and strong turbulence arguments both for isotropic and anisotropic fields \cite{biskamp03,zhouetal04,boldyrev06,ngbhattacharjee97,galtieretal00}. 
Following Pouquet et al. \cite{pouquetetal10}, the spectra for run TG are averaged between adjacent shells in order to get rid of the even-odd oscillations due to the specific structure of the TG configuration and obtain less biased plateaus in our spectra.

According to the simulations by Lee et al. \cite{leeetal10}, the total energy spectrum at the peak of dissipation for the insulating magnetic TG initial condition (Eqs. (4) and (5)) scales as $E_t(k) \propto k^{-2}$. The same scaling was also confirmed by our runs \cite{da13a} for the same initial conditions but without enforcing the TG symmetries. Looking at the spectra of the kinetic and magnetic energy individually, we observe that they exhibit different scalings exponents. In detail, Fig. \ref{fig:Eb_spectra}a shows the compensated magnetic energy spectrum for run TG, which scales very well like $E_b \propto k^{-2}$ for a decade of wavenumbers. These compensated spectra are clearly steeper for the $-5/3$ and $-3/2$ power laws, which can be excluded as possible fits to this spectrum. The compensated magnetic energy spectrum for run R (see Fig. \ref{fig:Eb_spectra}b) is clearly steeper than $k^{-2}$ and seems to scale as $E_b \propto k^{-5/3}$, whereas the kinetic energy  spectrum like $E_u \propto k^{-3/2}
$ (see Fig. \ref{fig:Eu_spectra}b) in agreement with previous works \cite{podestaetal07,alexakis13}. The difference between $k^{-5/3}$ and $k^{-3/2}$ power laws is subtle enough that any type of contamination, such as intermittency or any dissipative small-scale effects, will blur the results. Therefore, higher Reynolds number simulations would be required to have a clearer idea for these scalings in the inertial range. The compensated kinetic energy spectrum of run TG (see Fig. \ref{fig:Eu_spectra}a) is clearly steeper than the $k^{-5/3}$ spectrum and possibly even steeper than the $k^{-2}$ spectrum. However, we observe that the slope does not seem to be monotonic with two different peaks appearing in $k^pE_u$, one at large and one at smaller wavenumbers. Thus we do not have a clean power law behaviour for the energy spectrum at this Reynolds number. We denote that in run TG the total energy spectrum is dominated by $E_b$, since $E_b > E_u$  as one can also observe from Figs. \ref{fig:Eb_spectra} and \ref{
fig:Eu_spectra}. Thus, when plotting the total energy spectra of the TG flow it is the $k^{-2}$ behaviour of the magnetic field that is observed.

This scaling is in agreement with WT theory of Alfv\'en waves in the presence of a strong large scale magnetic field $B_0$. In this case, the sweeping effect of Alfv\'en waves propagation becomes important. Hence,
\begin{equation}
 \label{eq:wktime}
 \tau_A / \tau_{nl} \ll 1
\end{equation}
where $\tau_A \propto \ell_\parallel / B_0$ is the timescale associated with the propagation of Alfv\'en waves along the magnetic field lines of $B_0$ and $\tau_{nl} \propto \ell_\perp / u_\ell$ is the non-linear timescale related to the transfer of energy from an eddy of characteristic lengthscale and velocity to smaller eddies. If Eq. \eqref{eq:wktime} is valid, then turbulence is weak \cite{nazarenko11}, meaning that the non-linear energy transfer is delayed and thus the scaling of dissipation in the inertial range at high enough Reynolds numbers is taking the following form
\begin{equation}
 \label{eq:wkeps}
 \epsilon \propto \frac{u_\ell^2}{\tau_{nl}}\lt(\frac{\tau_A}{\tau_{nl}}\rt).
\end{equation}
Then, from Eq. \eqref{eq:wkeps} follows that the anisotropic weak turbulence energy spectrum is $E(k_\parallel,k_\perp) \propto f(k_\parallel) k_\perp^{-2}$ \cite{ngbhattacharjee97,galtieretal00}, whereas the Iroshnikov-Kraichnan spectrum is $E(k) \propto k^{-3/2}$ \cite{iroshnikov64,kraichnan65} assuming isotropy (i.e. $\ell_\perp \sim \ell_\parallel \sim \ell$).

Even though no external magnetic field was imposed in the present simulations, these particular initial conditions lead to the formation of large scale magnetic structures with the total magnetic energy growing significantly larger than the kinetic energy, i.e. $E_b \gtrsim 4 E_u$ (see also the rms values of $u'$ and $b'$ in Table \ref{tbl:dnsparam}, at the peak of dissipation). Hence for scales smaller than the integral scale $L$ it is assumed that the large scale magnetic field $B_L$ can be approximated as a
a quasi-uniform field for which Eq. \eqref{eq:wktime} applies and leads to the weak turbulence spectrum
observed in the simulations. It is thus tempting to interpret the $k^{-2}$ spectrum in terms of weak turbulence theory.

Although a strong magnetic field is a necessary condition for weak turbulence to occur, it is not sufficient for the energy spectrum to exhibit a $k^{-2}$ power law for various reasons, other than isotropy. Small scale variations $\ell_\perp$ can couple to large-scale parallel variations $\ell_\|$, with $\ell_\perp / u_\ell \propto \ell_\| / B_L$ so  that strong turbulence becomes important. In this case, the scaling of the dissipation rate is not expressed by Eq. \eqref{eq:wkeps} but it rather takes the classical form, i.e. $\epsilon \propto u^2_\ell / \tau_{nl}$ \cite{taylor35}. 
In addition, it was recently shown by Alexakis \cite{alexakis13} that Eq. \eqref{eq:wktime} is not necessarily valid for MHD turbulence with zero flux large-scale magnetic fields even when $E_b \gg E_u$. Thus, the weak turbulence scaling for the energy spectrum was absent from that investigation. As a result, based on the condition $E_b \gg E_u$ alone we cannot a priori decide if turbulence falls in the weak or strong turbulence regime. Therefore, we infer that $k^{-2}$ scaling for run TG could possibly manifest from rather different origins than weak interactions of Alfv\'en waves.

Further insight can be obtained by looking at the structures developed in the TG flow. These structures were analysed in \cite{da13a} by classifying their local topology. According to Dallas \& Alexakis \cite{da13a}, the dominant structures in run TG both in the vorticity and current density fields can be characterised as quasi-2D structures that are formed at the faces of the $[0,\pi]^3$ sub-boxes of the $[0,2\pi]^3$ periodic box. This type of quasi-2D structures were also observed in the current density field of run R (see \cite{da13a}), however, in this case the current sheets are weaker and randomly oriented in contrast to the structures of run TG, which are well organised. 
Thereafter, we investigate what is the influence of these quasi-2D current sheetlike structures on the magnetic energy spectrum.

\subsection{\label{sec:disc}The spectrum of discontinuities}
\begin{figure}[!ht]
  \begin{subfigure}{6cm}
%  \begin{subfigure}{5cm}
   \includegraphics[width=\textwidth]{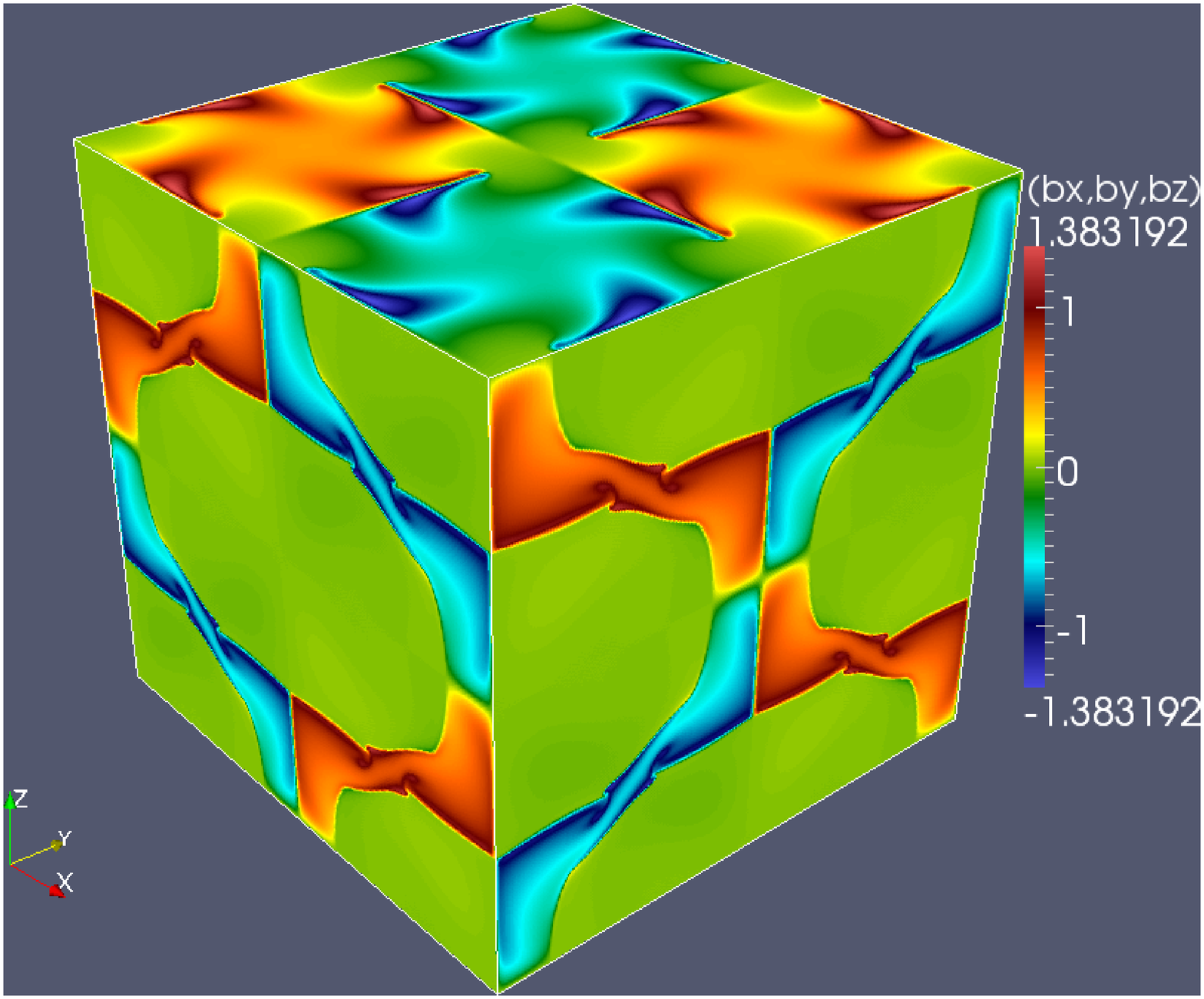}
  \caption{}
 \end{subfigure}
  \begin{subfigure}{6cm}
%  \begin{subfigure}{5.2cm}
   \includegraphics[width=\textwidth]{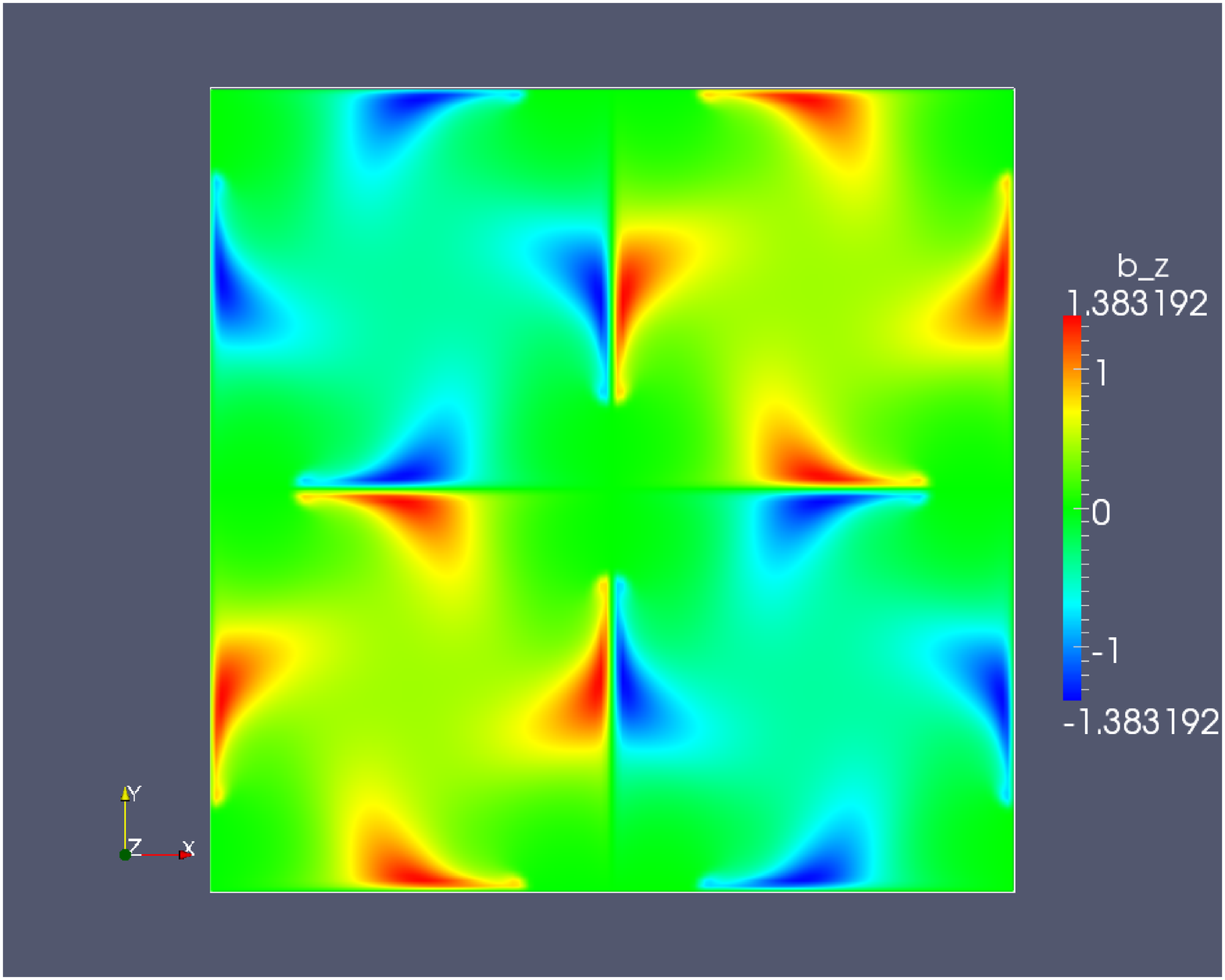}
  \caption{}
 \end{subfigure}
  \begin{subfigure}{6cm}
%  \begin{subfigure}{5.4cm}
   \includegraphics[width=\textwidth]{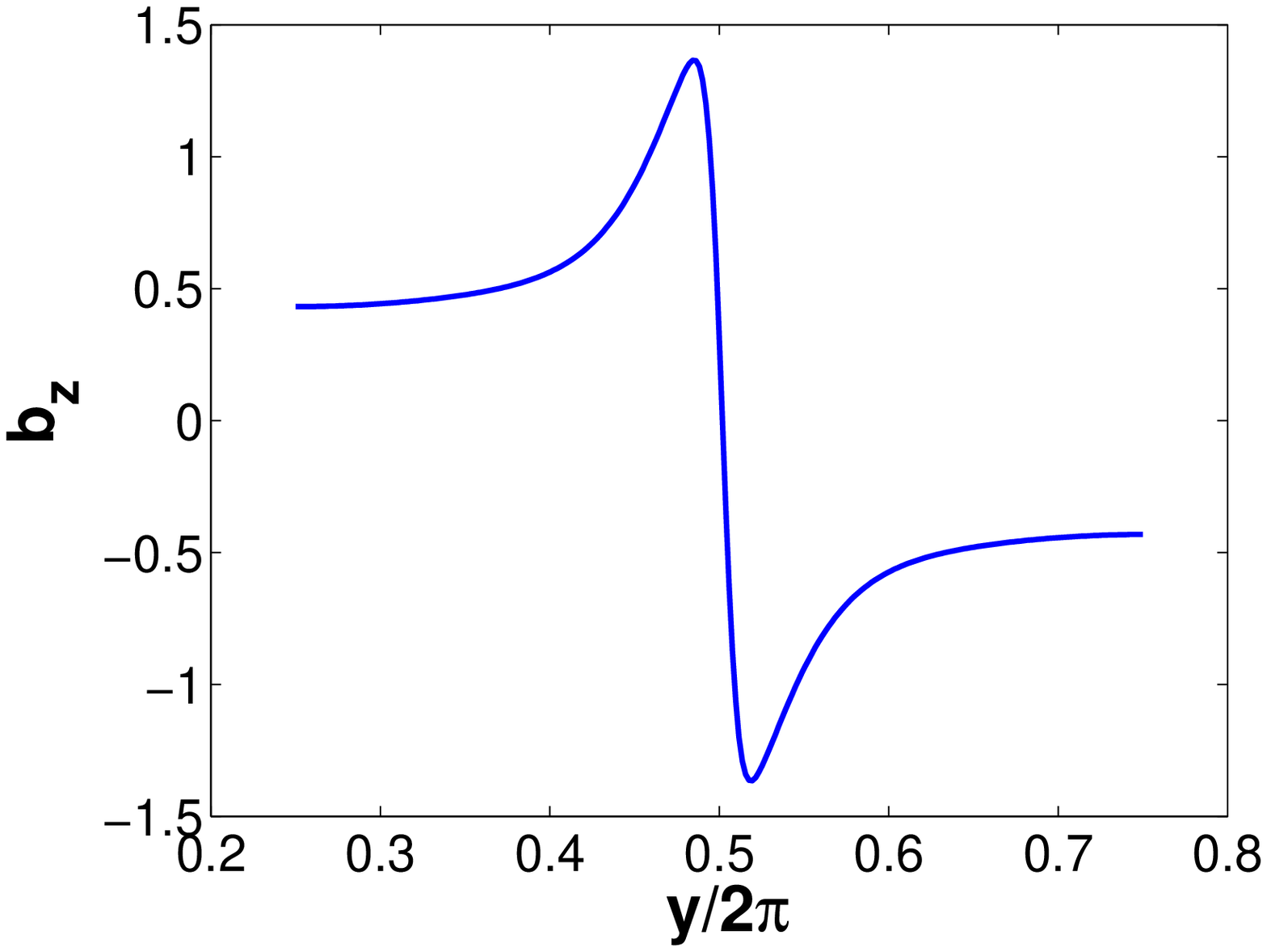}
  \caption{}
 \end{subfigure}
  \caption{(Color online) a) Contours of the individual components of the magnetic field $(b_x,b_y,b_z)$ at the peak of dissipation plotted on the corresponding $(x,y,z)$ faces of the periodic box. b) View of the $b_z$ component on the plane $z = 2\pi$. c) Profile of $b_z$ as a function of $y/2\pi$ at $x/2\pi = 0.2$.}
  \label{fig:bfield}
 \end{figure}

The quasi-2D structures observed in the current density of run TG are created due to strong shearing as it was mentioned in \cite{da13a}. This can be obvious when looking at the individual magnetic field components $\bm b = (b_x,b_y,b_z)$ at the peak of dissipation, which are presented in Fig. \ref{fig:bfield}a on the $(x,y,z)$ faces of the periodic box, respectively. One can easily notice that strong shear layers exist in the magnetic field with the red regions corresponding to positive values  (outwards from the box) and the blue regions to negative values (inwards to the box).
For clarity, we focus at the top face of the box (i.e. plane $z = 2\pi$), which shows the $b_z$ component (Fig. \ref{fig:bfield}b). To be more precise, we then plot the variation of $b_z$ in the $y$ direction at $x/2\pi = 0.2$, where we obtain a clear-cut high shearing profile (see Fig. \ref{fig:bfield}c), which could be represented by a Heaviside function $H(y)$. From there one finds that 
\begin{equation}
 j_x = \pd_y b_z = \delta(y)
\end{equation}
from definition, i.e. $H(y) \equiv \int_{-\infty}^y \delta(s) \, \dd s$, where $\delta$ is the Dirac delta function. The Fourier transform of a $\delta$ function in three dimensions gives 
\begin{equation}
 \hat{j}_x(\bm k) = \int_{-\infty}^{+\infty} \delta(y) e^{i \bm k \bm x} \, \dd^3 \bm x \propto \delta(k_x) \delta(k_z)
\end{equation}
since $\delta(y) \equiv \frac{1}{2\pi} \int_{-\infty}^{+\infty} e^{-iky} \, \dd k$, which is the definition of the $\delta$ function in terms of the Fourier integrals. The integration of $\hat{j}_x^2(\bm k)$ over all spherical cells in spectral space is
\begin{equation}
 \hat{j}_x^2(\bm k) \propto \int_k^{k+1} \delta(k'_x)\delta(k'_z) \, \dd^3 \bm k' = 1.
\end{equation}
So, the magnetic energy spectrum will then be
\begin{equation}
 E_b(k) = \frac{1}{2} \sum_k^{k+1} \frac{\hat{j}_x^2(\bm k')}{k'^2} \propto k^{-2}.
\end{equation}
Hence, the $k^{-2}$ spectrum observed in our run TG is due to extreme shearing regions that manifest discontinuities in the magnetic field corresponding to the quasi-2D structures of the current density. This result is in analogy with Burger's turbulence \cite{beckhanin07}, where a $k^{-2}$ scaling law for the energy spectrum also emerges due to discontinues in the velocity field.

It is well known that current sheets form spontaneously in MHD turbulence, providing a natural source of discontinuities \cite{parker94}. It is interesting that magnetic discontinuities have been recently observed in the solar wind \cite{grecoetal10} and it is hypothesised that 
they are generated predominantly by non-linear interactions \cite{li08}.

Nevertheless, one has to be careful when relating structures to spectra. A nice exposition of misleading examples can be found in \cite{tsinober09}. 
Therefore, in the next section we try to strengthen our argument by studying the influence of the structures on the energy spectrum.

\section{\label{sec:filter}Filtering}
In order to identify the role of the quasi-2D current sheets on $E_b$, we would like somehow to isolate these coherent structures from the background current density and check if the $k^{-2}$ spectrum precisely originates from these regions of strong shear or not. To do this we generate a new field $\bm b^>$ by eliminating the current density at points where $|\bm j| < j_{cut}$ (i.e. high pass filter) with $j_{cut}$ a given threshold. So, we set
\begin{equation}
\bm j_0 =
   \begin{cases}
    \bm j & \text{if } |\bm j| \ge j_{cut} \\
    0     & \text{otherwise}.
   \end{cases}
\end{equation}
and we make this field solenoidal by projection, i.e. 
\begin{equation}
 \bm j^> = \bm j_0 - \grad \phi
\end{equation}
where the scalar $\phi = \grad^{-2}(\grad \sdot \bm j_0)$. Then, the new field $\bm j^>$ is the solenoidal projection of $\bm j_0$ satisfying $\grad \sdot \bm j^> = 0$. Ultimately, the filtered magnetic field can be computed as
\begin{equation}
 \bm b^> = -\grad^{-2}(\grad\times \bm j^>)
\end{equation}
and its energy spectrum $E^>_b(k)$ can also be obtained. Note that the current density of $\bm j^>$ is not strictly zero outside the current sheets that we want to isolate but the above variational analysis guarantees that the residual $-\grad \phi$ is minimal by satisfying the Poisson equation (see also \cite{jimenezetal93}).

So, we apply this high pass filter to the current density, at the peak of dissipation, of run TG but also of run R so that we highlight the influence of the filtering in each case. The first column of Table \ref{tbl:hpfilter} lists the percentage cut off in terms of the maximum current density ($j_{cut}/j_{max}$), which is common for runs TG and R. The second and third columns represent essentially the percentage of Ohmic dissipation that is kept in the flow field after the filtering ($j^2_{kept}/j^2_{total}$) for run TG and R, respectively. Finally, the third and fourth columns show the volume percentage ($V_{kept}/V_{total}$) of the structures related to $j^2_{kept}$ for runs TG and R.
\begin{table}[!ht]
 \caption{Percentage of high pass filtering of the current density field.}
 \label{tbl:hpfilter}
  \begin{ruledtabular}
% \resizebox{0.5\textwidth}{!}
% {
     \begin{tabular}{c*{4}{c}}
%     \hline
%     \hline   
     $\bm{j_{cut}/j_{max}\; (\%)}$ & \multicolumn{2}{c}{$\bm{j^2_{kept}/j^2_{total} \; (\%)}$} & \multicolumn{2}{c}{$\bm{V_{kept}/V_{total}\; (\%)}$} \\
     Runs TG \& R & Run TG & Run R & Run TG & Run R \\
      \hline
      0  & 100 & 100 & 100 & 100 \\
      5  &  92 &  82 &   8 &  20 \\
      10 &  88 &  63 &   6 &   8 \\
      20 &  74 &  30 &   3 &   1 \\
      30 &  51 &  16 &   1 & 0.4 \\
%     \hline
%     \hline 
    \end{tabular}
% }
 \end{ruledtabular}
\end{table}

According to Table \ref{tbl:hpfilter}, for run TG with only 5\% cut-off of $j_{max}$, we keep
92\% of $j^2_{total}$ where this is concentrated at the utmost 8\% of the total volume of the box.
This 8\% of the volume corresponds to the structures at the faces of the $[0,\pi]^3$ sub-boxes where the discontinuities of the magnetic field can be seen in Fig. \ref{fig:bfield}a.
For run R, where current sheets are not so strong, a 5\% cut-off keeps 82\% of the total Ohmic dissipation in the flow, which is associated with structures that occupy 20\% of $V_{total}$ (see Table \ref{tbl:hpfilter}). Note that the $j_{rms}$ of the original field is 7\% and 6\% of the $j_{max}$ for runs TG and R, respectively. The different levels of filtering in Table \ref{tbl:hpfilter} delineate how diversely the Ohmic dissipation is distributed within the two flows that we deal with in this study.

\begin{figure}[!ht]
  \begin{subfigure}{6cm}
   \includegraphics[width=\textwidth]{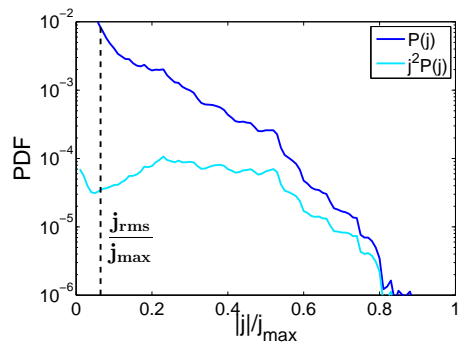}
   \caption{}
  \end{subfigure}
  \begin{subfigure}{6cm}
   \includegraphics[width=\textwidth]{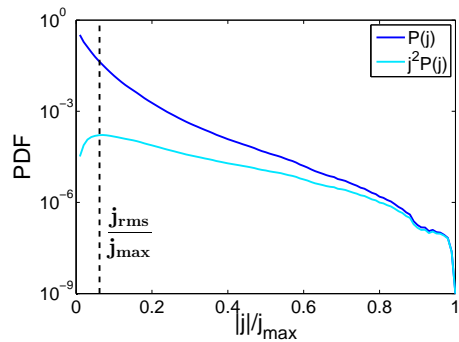}
   \caption{}
  \end{subfigure}
  \caption{(Color online) The PDF of $|\bm j|/j_{max}$ and the curve of $j^2P(j)$ appropriately normalised with $j_{max}$ for (a) run TG and (b) run R. Note that $j_{rms}/j_{max} = 0.07$ and 0.06, respectively.}
  \label{fig:pdf}
 \end{figure}

The probability density function (PDF) of $|\bm j|/j_{max}$ for runs TG and R are presented in Figs. \ref{fig:pdf}a and \ref{fig:pdf}b, respectively. In order to indicate where most of the Ohmic dissipation occurs, we note that $\avg{|\bm j|^2} = \int_{-\infty}^{+\infty} j^2P(j) dj$ and we include in Fig. \ref{fig:pdf} the curve of $j^2P(j)$ normalised appropriately with $j_{max}$. Therefore, we see that most of the dissipation in run R occurs around $j_{rms}/j_{max} = 6\%$ with monotonic drop of the PDF for larger values (see Fig. \ref{fig:pdf}b). 
The PDF of run TG is significantly different with the maximum $j^2P(j)$ occurring at much higher values than $j_{rms}/j_{max} = 7\%$ (see dashed line in Fig. \ref{fig:pdf}a), i.e. $20\% \lesssim |\bm j|/j_{max} \lesssim 40\%$, indicating that a few points of the computational box give significant contribution to the total Ohmic dissipation. It is worth noting that for $\bm j^>$ with $j_{cut}/j_{max} = 5\%$ we keep all these extreme points but we throw away the left part from the dashed line.

\begin{figure}[!ht]
  \begin{subfigure}{6cm}
   \includegraphics[width=\textwidth]{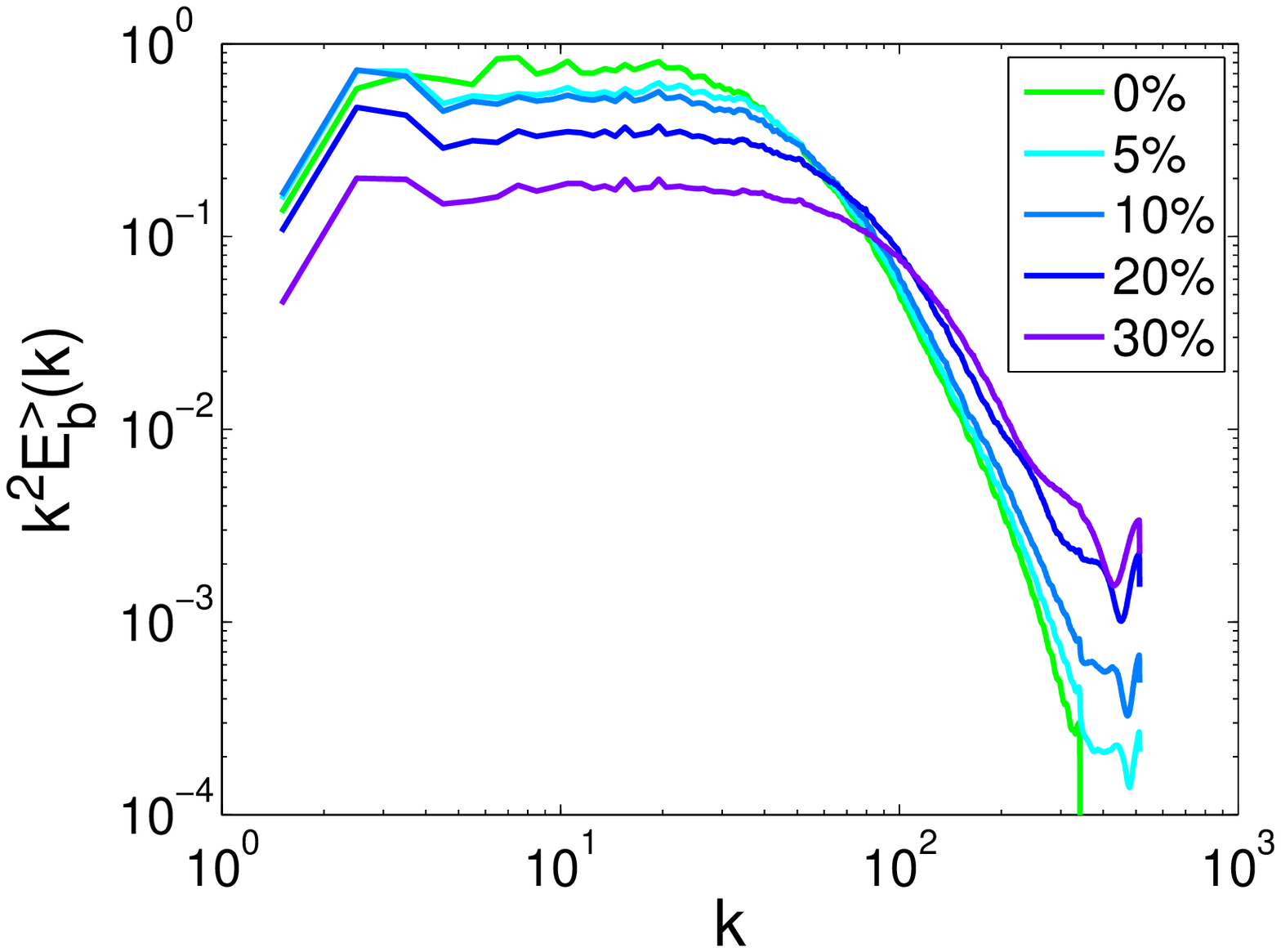}
   \caption{}
  \end{subfigure}
  \begin{subfigure}{6cm}
    \includegraphics[width=\textwidth]{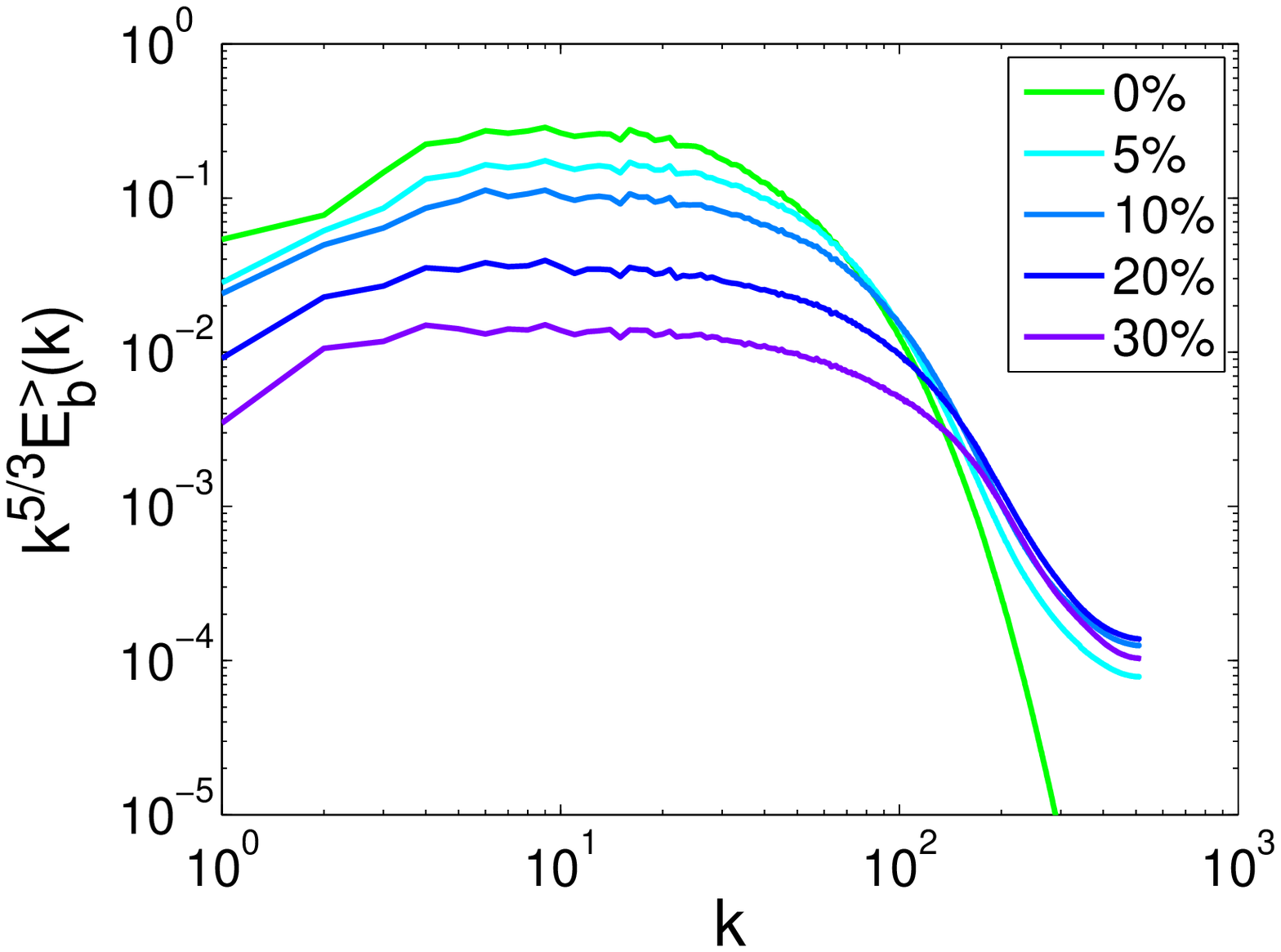}
   \caption{}
  \end{subfigure}
   \caption{(Color online) The high pass filtered magnetic energy spectrum $E^>_b$ compensated with $k^{2}$ at various levels of filtering for (a) run TG and (b) run R.}
   \label{fig:LF_spectra}
  \end{figure}

Figure \ref{fig:LF_spectra} presents the energy spectra $E_b^>$ compensated with $k^p$ where $p=2$ and $5/3$ for runs TG and R, respectively.
Increasing gradually the cut-off threshold of the filtering, we can observe that the high wavenumber end of the spectrum $E^>_b$ is modified due to sharp filter. It is clearly demonstrated in Fig. \ref{fig:LF_spectra}a that the original $k^{-2}$ scaling is not affected by the filtering in an intermediate range $\eta \ll \ell \ll L$. Thus, we deduce that the $-2$ power law in run TG can be reconstructed just from the structures that occupy only 1\% of the total volume and accommodate almost half ($\simeq 51\%$) of the Ohmic dissipation if we consider the extreme case of 30\% filtering of $j_{max}$. This outcome supports our argument that $E_b \propto k^{-2}$ originates from the regions of strong magnetic shear with discontinuous profile (see Fig. \ref{fig:bfield}), where quasi-2D structures are formed. Similarly, the scaling of the $E^>_b$ spectrum of run R (see Fig. \ref{fig:LF_spectra}b) does not deviate from the original $k^{-5/3}$ in an intermediate range $\eta \ll \ell \ll L$, even if we cut off 30\% of 
$j_{max}$ and we are left with the structures that hold only 0.4\% of $V_{total}$ to which attribute the 16\% of $j^2_{total}$. This result is in analogy to studies in hydrodynamic turbulence using several eduction techniques such the Karhunen-Lo\'eve decomposition \cite{holmesetal98}, where only a few modes are necessary to reconstruct various statistics of the turbulent flows.

One could still argue that weak Alfv\'en wave interactions, emanating away from the regions where discontinuities appear in the magnetic field, could potentially contribute to the formation of the $-2$ power law in the $E_b$ spectrum. To counteract on this argument, we also filter out the current density field at points where $|\bm j| > j_{cut}$ (i.e. low pass filter). We then compute $\bm j^<$ as well as its spectrum $E^<_b(k)$, where the quasi-2D current sheets from the strong shearing regions are truncated, so that we scrutinize if the $k^{-2}$ scaling remains in an intermediate range of scales. Here, we should mention that the values of $j^2_{kept}/j^2_{total}$ and $V_{kept}/V_{total}$ for the low pass filter is simply what is left from the high pass filter if we subtract the total amounts accordingly (see Table \ref{tbl:hpfilter}).
 \begin{figure}[!ht]
  \begin{subfigure}{6cm}
   \includegraphics[width=\textwidth]{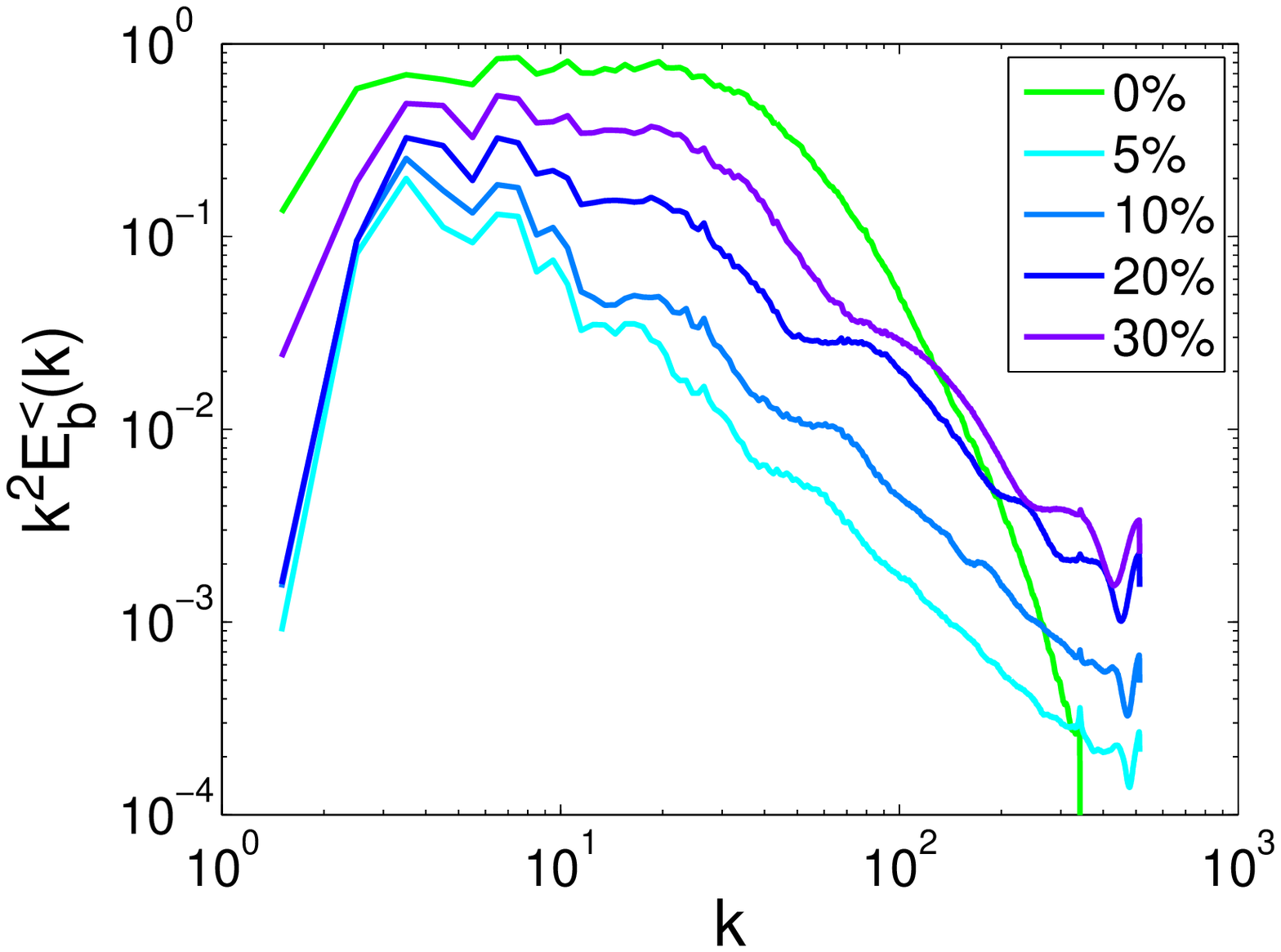}
   \caption{}
  \end{subfigure}
  \begin{subfigure}{6cm}
    \includegraphics[width=\textwidth]{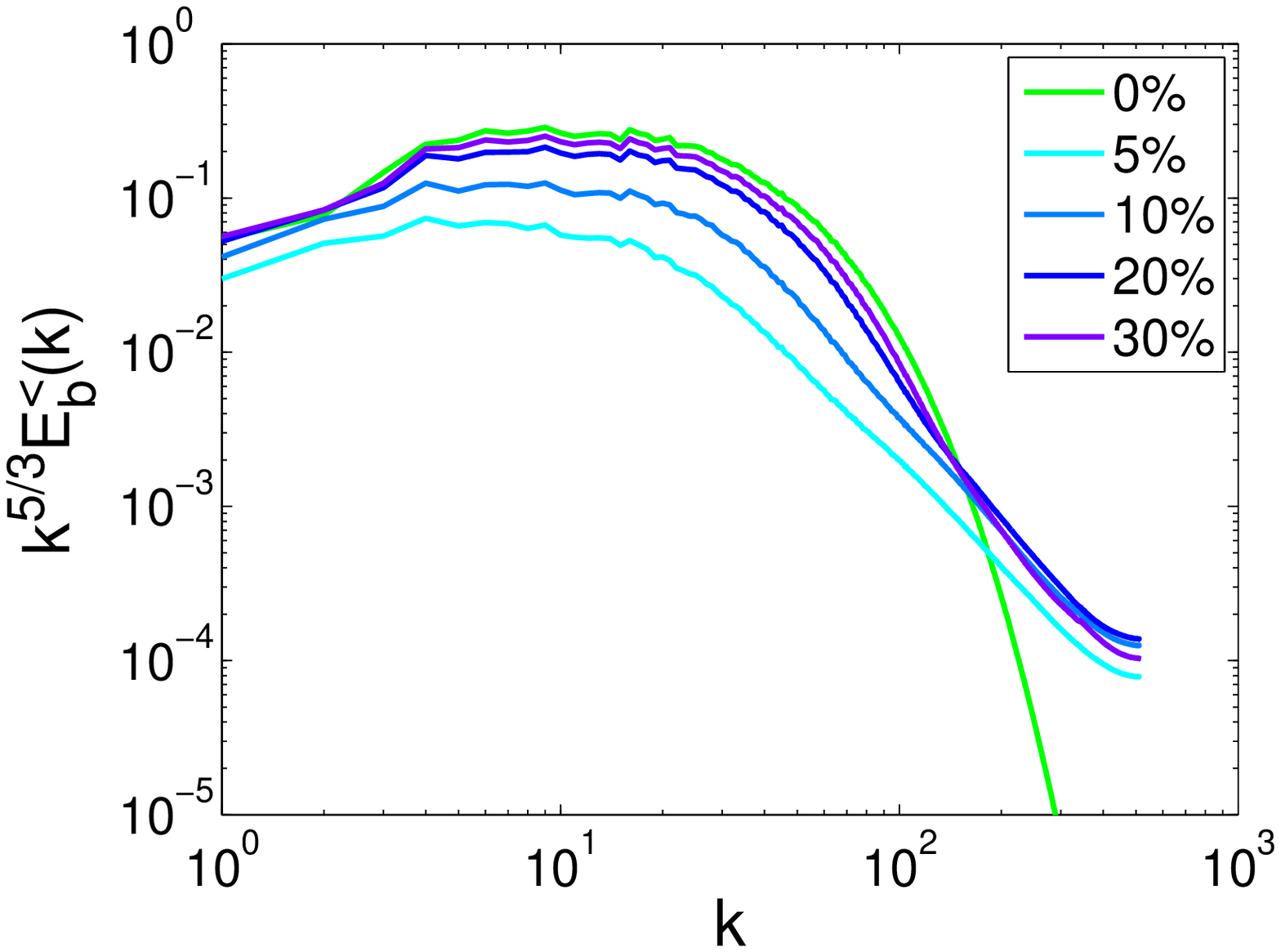}
   \caption{}
  \end{subfigure}
  \caption{(Color online) The low pass filtered magnetic energy spectrum $E^<_b$ compensated with $k^{5/3}$ at various levels of filtering for (a) run TG and (b) run R.}
  \label{fig:HF_spectra}
 \end{figure}

So, with reference to Table \ref{tbl:hpfilter} and Fig. \ref{fig:HF_spectra}a, we deduce that for $j_{cut}/j_{max} = 5\%$, where we keep only 8\% of $j^2_{total}$ but almost the whole volume (i.e. 92\% of $V_{total}$) occupied by the remaining structures in the field, the compensated spectrum $E^<_b$ is not even nearly to a $k^{-2}$ spectrum. This clearly demonstrates that the $-2$ power law is due to the strong current sheets that occupy only the 8\% of the volume of the computational box and concentrate the 92\% of the Ohmic dissipation of the flow (see Table \ref{tbl:hpfilter} and Fig. \ref{fig:LF_spectra}a). Of course, as we increase the cut-off threshold, elements from the high shear regions are involved and therefore we slowly start to recover the original spectrum. On the other side, the scaling of $E^<_b$ for run R does not go that far from the original $k^{-5/3}$ spectrum, which is recovered much faster than for run TG as we increase $j_{cut}/j_{max}$ (see Fig. \ref{fig:HF_spectra}b).

\section{\label{sec:end}Conclusions}
In the presence of a strong mean magnetic field, it is assumed that $\tau_A \ll \tau_{nl}$ and according to weak turbulence theory an anisotropic energy spectrum scales as $k_\perp^{-2}$. A paper by Lee et al. \cite{leeetal10} obtained $k^{-2}$, $k^{-5/3}$ and $k^{-3/2}$ scalings for different initial conditions of the magnetic field, showing dependence of the energy spectrum at the peak of dissipation on the initial conditions. It has been hypothesised in \cite{leeetal10} that weak turbulence phenomenology is a possible candidate to explain the $k^{-2}$ scaling in their total energy spectrum. However, their DNS have zero flux and their flows are fully non-linear composed by quasi-2D vortex and current sheets \cite{da13a}.

In this paper, we replicate the insulating magnetic TG initial condition from \cite{leeetal10} without enforcing the TG symmetries and we also obtain a clear $-2$ power law for the total energy spectrum (see also \cite{da13a}). We further observe that this law for the total energy emerges due to the scaling of the magnetic energy spectrum, since $E_b > E_v$ at the peak of dissipation. Then, looking in more detail at the magnetic field of our DNS, we observe that the quasi-2D current sheets are created in regions of strong shear, where the magnetic field changes direction abruptly and therefore forms discontinues profiles. Using this result, we are able to derive analytically the spectrum of discontinuities in the magnetic field, which entails a $k^{-2}$ scaling, demonstrating the origin of this scaling exponent from the numerical simulations.

To strengthen our claim, we study the effect of the quasi-2D current sheets on the $E_b$ spectra by isolating/eliminating the regions with strong shear in the current density field from the rest of the flow using a filtering technique. From there, we can clearly observe that the $-2$ power law actually emerges due to the regions that manifest discontinuities in the magnetic field and not due to any other turbulent effects.

The presence of a clear $k^{-2}$ spectrum due to the strong current sheets implies lack of universality in decaying MHD turbulence. However, an important point regarding the TG flows, that has to be addressed before claiming non-universality, is the role of the TG symmetries imposed by the initial conditions and their self-preservation in time evolution of the flow. In other words, are these discontinuities formed due to TG symmetries or are there more random cases where a $-2$ spectrum emerges due to discontinuities? What happens if we somehow break the TG symmetries before the peak of dissipation? Will the scaling of the energy spectra converge to a single value? Do we have classes of universality for these moderate Reynolds numbers or is there a universal power law for high Reynolds number limit? These are questions we plan to address in our future work.

\begin{acknowledgements}
The authors acknowledge stimulating discussions with Christos Vassilicos and Marc-Etienne Brachet. V.D. acknowledges the financial support from EU-funded Marie Curie Actions---Intra-European Fellowships (FP7-PEOPLE-2011-IEF, MHDTURB, Project No. 299973). The computations were performed using the HPC resources from GENCI-CINES-JADE (Project No. 2012026421), GENCI-TGCC-CURIE (Project No. x2013056421) and PRACE-FZJ-JUQUEEN (Project name PRA068).
\end{acknowledgements}
\bibliography{references}

\begin{thebibliography}{32}%
\makeatletter
\providecommand \@ifxundefined [1]{%
 \@ifx{#1\undefined}
}%
\providecommand \@ifnum [1]{%
 \ifnum #1\expandafter \@firstoftwo
 \else \expandafter \@secondoftwo
 \fi
}%
\providecommand \@ifx [1]{%
 \ifx #1\expandafter \@firstoftwo
 \else \expandafter \@secondoftwo
 \fi
}%
\providecommand \natexlab [1]{#1}%
\providecommand \enquote  [1]{``#1''}%
\providecommand \bibnamefont  [1]{#1}%
\providecommand \bibfnamefont [1]{#1}%
\providecommand \citenamefont [1]{#1}%
\providecommand \href@noop [0]{\@secondoftwo}%
\providecommand \href [0]{\begingroup \@sanitize@url \@href}%
\providecommand \@href[1]{\@@startlink{#1}\@@href}%
\providecommand \@@href[1]{\endgroup#1\@@endlink}%
\providecommand \@sanitize@url [0]{\catcode `\\12\catcode `\$12\catcode
  `\&12\catcode `\#12\catcode `\^12\catcode `\_12\catcode `\%12\relax}%
\providecommand \@@startlink[1]{}%
\providecommand \@@endlink[0]{}%
\providecommand \url  [0]{\begingroup\@sanitize@url \@url }%
\providecommand \@url [1]{\endgroup\@href {#1}{\urlprefix }}%
\providecommand \urlprefix  [0]{URL }%
\providecommand \Eprint [0]{\href }%
\providecommand \doibase [0]{http://dx.doi.org/}%
\providecommand \selectlanguage [0]{\@gobble}%
\providecommand \bibinfo  [0]{\@secondoftwo}%
\providecommand \bibfield  [0]{\@secondoftwo}%
\providecommand \translation [1]{[#1]}%
\providecommand \BibitemOpen [0]{}%
\providecommand \bibitemStop [0]{}%
\providecommand \bibitemNoStop [0]{.\EOS\space}%
\providecommand \EOS [0]{\spacefactor3000\relax}%
\providecommand \BibitemShut  [1]{\csname bibitem#1\endcsname}%
\let\auto@bib@innerbib\@empty
%</preamble>
\bibitem [{\citenamefont {Lee}\ \emph {et~al.}(2010)\citenamefont {Lee},
  \citenamefont {Brachet}, \citenamefont {Pouquet}, \citenamefont {Mininni},\
  and\ \citenamefont {Rosenberg}}]{leeetal10}%
  \BibitemOpen
  \bibfield  {author} {\bibinfo {author} {\bibfnamefont {E.}~\bibnamefont
  {Lee}}, \bibinfo {author} {\bibfnamefont {M.~E.}\ \bibnamefont {Brachet}},
  \bibinfo {author} {\bibfnamefont {A.}~\bibnamefont {Pouquet}}, \bibinfo
  {author} {\bibfnamefont {P.~D.}\ \bibnamefont {Mininni}}, \ and\ \bibinfo
  {author} {\bibfnamefont {D.}~\bibnamefont {Rosenberg}},\ }\href@noop {}
  {\bibfield  {journal} {\bibinfo  {journal} {Phys. Rev. E}\ }\textbf {\bibinfo
  {volume} {81}},\ \bibinfo {pages} {016318} (\bibinfo {year}
  {2010})}\BibitemShut {NoStop}%
\bibitem [{\citenamefont {Biskamp}(2003)}]{biskamp03}%
  \BibitemOpen
  \bibfield  {author} {\bibinfo {author} {\bibfnamefont {D.}~\bibnamefont
  {Biskamp}},\ }\href@noop {} {\emph {\bibinfo {title} {Magnetohydrodynamic
  turbulence}}}\ (\bibinfo  {publisher} {Cambridge University Press},\ \bibinfo
  {year} {2003})\BibitemShut {NoStop}%
\bibitem [{\citenamefont {Zhou}\ \emph {et~al.}(2004)\citenamefont {Zhou},
  \citenamefont {Matthaeus},\ and\ \citenamefont {Dmitruk}}]{zhouetal04}%
  \BibitemOpen
  \bibfield  {author} {\bibinfo {author} {\bibfnamefont {Y.}~\bibnamefont
  {Zhou}}, \bibinfo {author} {\bibfnamefont {W.~H.}\ \bibnamefont {Matthaeus}},
  \ and\ \bibinfo {author} {\bibfnamefont {P.}~\bibnamefont {Dmitruk}},\ }\href
  {\doibase 10.1103/RevModPhys.76.1015} {\bibfield  {journal} {\bibinfo
  {journal} {Rev. Mod. Phys.}\ }\textbf {\bibinfo {volume} {76}},\ \bibinfo
  {pages} {1015} (\bibinfo {year} {2004})}\BibitemShut {NoStop}%
\bibitem [{\citenamefont {Boldyrev}(2006)}]{boldyrev06}%
  \BibitemOpen
  \bibfield  {author} {\bibinfo {author} {\bibfnamefont {S.}~\bibnamefont
  {Boldyrev}},\ }\href@noop {} {\bibfield  {journal} {\bibinfo  {journal}
  {Phys. Rev. Lett.}\ }\textbf {\bibinfo {volume} {96}},\ \bibinfo {pages}
  {115002} (\bibinfo {year} {2006})}\BibitemShut {NoStop}%
\bibitem [{\citenamefont {Marino}\ \emph {et~al.}(2008)\citenamefont {Marino},
  \citenamefont {Sorriso-Valvo}, \citenamefont {Carbone}, \citenamefont
  {Noullez}, \citenamefont {Bruno},\ and\ \citenamefont
  {Bavassano}}]{marinoetal08}%
  \BibitemOpen
  \bibfield  {author} {\bibinfo {author} {\bibfnamefont {R.}~\bibnamefont
  {Marino}}, \bibinfo {author} {\bibfnamefont {L.}~\bibnamefont
  {Sorriso-Valvo}}, \bibinfo {author} {\bibfnamefont {V.}~\bibnamefont
  {Carbone}}, \bibinfo {author} {\bibfnamefont {A.}~\bibnamefont {Noullez}},
  \bibinfo {author} {\bibfnamefont {R.}~\bibnamefont {Bruno}}, \ and\ \bibinfo
  {author} {\bibfnamefont {B.}~\bibnamefont {Bavassano}},\ }\href@noop {}
  {\bibfield  {journal} {\bibinfo  {journal} {The Astrophysical Journal
  Letters}\ }\textbf {\bibinfo {volume} {677}},\ \bibinfo {pages} {L71}
  (\bibinfo {year} {2008})}\BibitemShut {NoStop}%
\bibitem [{\citenamefont {M{\"u}ller}\ and\ \citenamefont
  {Grappin}(2005)}]{mullergrappin05}%
  \BibitemOpen
  \bibfield  {author} {\bibinfo {author} {\bibfnamefont {W.~C.}\ \bibnamefont
  {M{\"u}ller}}\ and\ \bibinfo {author} {\bibfnamefont {R.}~\bibnamefont
  {Grappin}},\ }\href@noop {} {\bibfield  {journal} {\bibinfo  {journal} {Phys.
  Rev. Lett.}\ }\textbf {\bibinfo {volume} {95}},\ \bibinfo {pages} {114502}
  (\bibinfo {year} {2005})}\BibitemShut {NoStop}%
\bibitem [{\citenamefont {Mininni}\ and\ \citenamefont
  {Pouquet}(2007)}]{mininnipouquet07}%
  \BibitemOpen
  \bibfield  {author} {\bibinfo {author} {\bibfnamefont {P.~D.}\ \bibnamefont
  {Mininni}}\ and\ \bibinfo {author} {\bibfnamefont {A.}~\bibnamefont
  {Pouquet}},\ }\href@noop {} {\bibfield  {journal} {\bibinfo  {journal} {Phys.
  Rev. Lett.}\ }\textbf {\bibinfo {volume} {99}},\ \bibinfo {pages} {254502}
  (\bibinfo {year} {2007})}\BibitemShut {NoStop}%
\bibitem [{\citenamefont {Podesta}\ \emph {et~al.}(2007)\citenamefont
  {Podesta}, \citenamefont {Roberts},\ and\ \citenamefont
  {Goldstein}}]{podestaetal07}%
  \BibitemOpen
  \bibfield  {author} {\bibinfo {author} {\bibfnamefont {J.~J.}\ \bibnamefont
  {Podesta}}, \bibinfo {author} {\bibfnamefont {D.~A.}\ \bibnamefont
  {Roberts}}, \ and\ \bibinfo {author} {\bibfnamefont {M.~L.}\ \bibnamefont
  {Goldstein}},\ }\href@noop {} {\bibfield  {journal} {\bibinfo  {journal} {The
  Astrophysical Journal}\ }\textbf {\bibinfo {volume} {664}},\ \bibinfo {pages}
  {543} (\bibinfo {year} {2007})}\BibitemShut {NoStop}%
\bibitem [{\citenamefont {{J. Saur}}\ \emph {et~al.}(2002)\citenamefont {{J.
  Saur}}, \citenamefont {{H. Politano}}, \citenamefont {{A. Pouquet}},\ and\
  \citenamefont {{W. H. Matthaeus}}}]{sauretal02}%
  \BibitemOpen
  \bibfield  {author} {\bibinfo {author} {\bibnamefont {{J. Saur}}}, \bibinfo
  {author} {\bibnamefont {{H. Politano}}}, \bibinfo {author} {\bibnamefont {{A.
  Pouquet}}}, \ and\ \bibinfo {author} {\bibnamefont {{W. H. Matthaeus}}},\
  }\href {\doibase 10.1051/0004-6361:20020305} {\bibfield  {journal} {\bibinfo
  {journal} {Astron. Astrophys.}\ }\textbf {\bibinfo {volume} {386}},\ \bibinfo
  {pages} {699} (\bibinfo {year} {2002})}\BibitemShut {NoStop}%
\bibitem [{\citenamefont {Ng}\ and\ \citenamefont
  {Bhattacharjee}(1997)}]{ngbhattacharjee97}%
  \BibitemOpen
  \bibfield  {author} {\bibinfo {author} {\bibfnamefont {C.~S.}\ \bibnamefont
  {Ng}}\ and\ \bibinfo {author} {\bibfnamefont {A.}~\bibnamefont
  {Bhattacharjee}},\ }\href {\doibase 10.1063/1.872158} {\bibfield  {journal}
  {\bibinfo  {journal} {Phys. Plasmas}\ }\textbf {\bibinfo {volume} {4}},\
  \bibinfo {pages} {605} (\bibinfo {year} {1997})}\BibitemShut {NoStop}%
\bibitem [{\citenamefont {Galtier}\ \emph {et~al.}(2000)\citenamefont
  {Galtier}, \citenamefont {Nazarenko}, \citenamefont {Newell},\ and\
  \citenamefont {Pouquet}}]{galtieretal00}%
  \BibitemOpen
  \bibfield  {author} {\bibinfo {author} {\bibfnamefont {S.}~\bibnamefont
  {Galtier}}, \bibinfo {author} {\bibfnamefont {S.~V.}\ \bibnamefont
  {Nazarenko}}, \bibinfo {author} {\bibfnamefont {A.~C.}\ \bibnamefont
  {Newell}}, \ and\ \bibinfo {author} {\bibfnamefont {A.}~\bibnamefont
  {Pouquet}},\ }\href@noop {} {\bibfield  {journal} {\bibinfo  {journal} {J.
  Plasma Phys.}\ }\textbf {\bibinfo {volume} {63}},\ \bibinfo {pages} {447}
  (\bibinfo {year} {2000})}\BibitemShut {NoStop}%
\bibitem [{\citenamefont {Gottlieb}\ and\ \citenamefont
  {Orszag}(1977)}]{gottlieborszag77}%
  \BibitemOpen
  \bibfield  {author} {\bibinfo {author} {\bibfnamefont {D.}~\bibnamefont
  {Gottlieb}}\ and\ \bibinfo {author} {\bibfnamefont {S.}~\bibnamefont
  {Orszag}},\ }\href@noop {} {\emph {\bibinfo {title} {Numerical analysis of
  spectral methods: theory and applications}}},\ Vol.~\bibinfo {volume} {26}\
  (\bibinfo  {publisher} {SIAM},\ \bibinfo {year} {1977})\BibitemShut {NoStop}%
\bibitem [{\citenamefont {Frigo}\ and\ \citenamefont {Johnson}(1998)}]{fftw98}%
  \BibitemOpen
  \bibfield  {author} {\bibinfo {author} {\bibfnamefont {M.}~\bibnamefont
  {Frigo}}\ and\ \bibinfo {author} {\bibfnamefont {S.}~\bibnamefont
  {Johnson}},\ }in\ \href@noop {} {\emph {\bibinfo {booktitle} {Acoustics,
  Speech and Signal Processing, 1998. Proceedings of the 1998 IEEE
  International Conference on}}},\ Vol.~\bibinfo {volume} {3}\ (\bibinfo
  {organization} {IEEE},\ \bibinfo {year} {1998})\ pp.\ \bibinfo {pages}
  {1381--1384}\BibitemShut {NoStop}%
\bibitem [{\citenamefont {G\'omez}\ \emph {et~al.}(2005)\citenamefont
  {G\'omez}, \citenamefont {Mininni},\ and\ \citenamefont
  {Dmitruk}}]{mpicode05a}%
  \BibitemOpen
  \bibfield  {author} {\bibinfo {author} {\bibfnamefont {D.~O.}\ \bibnamefont
  {G\'omez}}, \bibinfo {author} {\bibfnamefont {P.~D.}\ \bibnamefont
  {Mininni}}, \ and\ \bibinfo {author} {\bibfnamefont {P.}~\bibnamefont
  {Dmitruk}},\ }\href {\doibase 10.1016/j.asr.2005.02.099} {\bibfield
  {journal} {\bibinfo  {journal} {Advances in Space Research}\ }\textbf
  {\bibinfo {volume} {35}},\ \bibinfo {pages} {899 } (\bibinfo {year}
  {2005})}\BibitemShut {NoStop}%
\bibitem [{\citenamefont {Mininni}\ \emph {et~al.}(2011)\citenamefont
  {Mininni}, \citenamefont {Rosenberg}, \citenamefont {Reddy},\ and\
  \citenamefont {Pouquet}}]{hybridcode11}%
  \BibitemOpen
  \bibfield  {author} {\bibinfo {author} {\bibfnamefont {P.~D.}\ \bibnamefont
  {Mininni}}, \bibinfo {author} {\bibfnamefont {D.}~\bibnamefont {Rosenberg}},
  \bibinfo {author} {\bibfnamefont {R.}~\bibnamefont {Reddy}}, \ and\ \bibinfo
  {author} {\bibfnamefont {A.}~\bibnamefont {Pouquet}},\ }\href {\doibase
  10.1016/j.parco.2011.05.004} {\bibfield  {journal} {\bibinfo  {journal}
  {Parallel Computing}\ }\textbf {\bibinfo {volume} {37}},\ \bibinfo {pages}
  {316 } (\bibinfo {year} {2011})}\BibitemShut {NoStop}%
\bibitem [{\citenamefont {Taylor}\ and\ \citenamefont
  {Green}(1937)}]{taylorgreen37}%
  \BibitemOpen
  \bibfield  {author} {\bibinfo {author} {\bibfnamefont {G.}~\bibnamefont
  {Taylor}}\ and\ \bibinfo {author} {\bibfnamefont {A.}~\bibnamefont {Green}},\
  }\href@noop {} {\bibfield  {journal} {\bibinfo  {journal} {Proc. R. Soc.
  London A}\ }\textbf {\bibinfo {volume} {158}},\ \bibinfo {pages} {499}
  (\bibinfo {year} {1937})}\BibitemShut {NoStop}%
\bibitem [{\citenamefont {Brachet}\ \emph {et~al.}(1983)\citenamefont
  {Brachet}, \citenamefont {Meiron}, \citenamefont {Orszag}, \citenamefont
  {Nickel}, \citenamefont {Morf},\ and\ \citenamefont
  {Frisch}}]{brachetetal83}%
  \BibitemOpen
  \bibfield  {author} {\bibinfo {author} {\bibfnamefont {M.~E.}\ \bibnamefont
  {Brachet}}, \bibinfo {author} {\bibfnamefont {D.~I.}\ \bibnamefont {Meiron}},
  \bibinfo {author} {\bibfnamefont {S.~A.}\ \bibnamefont {Orszag}}, \bibinfo
  {author} {\bibfnamefont {B.~G.}\ \bibnamefont {Nickel}}, \bibinfo {author}
  {\bibfnamefont {R.~H.}\ \bibnamefont {Morf}}, \ and\ \bibinfo {author}
  {\bibfnamefont {U.}~\bibnamefont {Frisch}},\ }\href {\doibase
  10.1017/S0022112083001159} {\bibfield  {journal} {\bibinfo  {journal} {J.
  Fluid Mech.}\ }\textbf {\bibinfo {volume} {130}},\ \bibinfo {pages} {411}
  (\bibinfo {year} {1983})}\BibitemShut {NoStop}%
\bibitem [{\citenamefont {Dallas}\ and\ \citenamefont
  {Alexakis}(2013)}]{da13a}%
  \BibitemOpen
  \bibfield  {author} {\bibinfo {author} {\bibfnamefont {V.}~\bibnamefont
  {Dallas}}\ and\ \bibinfo {author} {\bibfnamefont {A.}~\bibnamefont
  {Alexakis}},\ }\href {http://arxiv.org/abs/1304.0695} {\bibfield  {journal}
  {\bibinfo  {journal} {submitted in Phys. Fluids,
  http://arxiv.org/abs/1304.0695}\ } (\bibinfo {year} {2013})}\BibitemShut
  {NoStop}%
\bibitem [{\citenamefont {Lee}\ \emph {et~al.}(2008)\citenamefont {Lee},
  \citenamefont {Brachet}, \citenamefont {Pouquet}, \citenamefont {Mininni},\
  and\ \citenamefont {Rosenberg}}]{leeetal08}%
  \BibitemOpen
  \bibfield  {author} {\bibinfo {author} {\bibfnamefont {E.}~\bibnamefont
  {Lee}}, \bibinfo {author} {\bibfnamefont {M.~E.}\ \bibnamefont {Brachet}},
  \bibinfo {author} {\bibfnamefont {A.}~\bibnamefont {Pouquet}}, \bibinfo
  {author} {\bibfnamefont {P.~D.}\ \bibnamefont {Mininni}}, \ and\ \bibinfo
  {author} {\bibfnamefont {D.}~\bibnamefont {Rosenberg}},\ }\href {\doibase
  10.1103/PhysRevE.78.066401} {\bibfield  {journal} {\bibinfo  {journal} {Phys.
  Rev. E}\ }\textbf {\bibinfo {volume} {78}},\ \bibinfo {pages} {066401}
  (\bibinfo {year} {2008})}\BibitemShut {NoStop}%
\bibitem [{\citenamefont {Pouquet}\ \emph {et~al.}(2010)\citenamefont
  {Pouquet}, \citenamefont {Lee}, \citenamefont {Brachet}, \citenamefont
  {Mininni},\ and\ \citenamefont {Rosenberg}}]{pouquetetal10}%
  \BibitemOpen
  \bibfield  {author} {\bibinfo {author} {\bibfnamefont {A.}~\bibnamefont
  {Pouquet}}, \bibinfo {author} {\bibfnamefont {E.}~\bibnamefont {Lee}},
  \bibinfo {author} {\bibfnamefont {M.}~\bibnamefont {Brachet}}, \bibinfo
  {author} {\bibfnamefont {P.}~\bibnamefont {Mininni}}, \ and\ \bibinfo
  {author} {\bibfnamefont {D.}~\bibnamefont {Rosenberg}},\ }\href@noop {}
  {\bibfield  {journal} {\bibinfo  {journal} {Geophys. Astrophys. Fluid Dyn.}\
  }\textbf {\bibinfo {volume} {104}},\ \bibinfo {pages} {115} (\bibinfo {year}
  {2010})}\BibitemShut {NoStop}%
\bibitem [{\citenamefont {Alexakis}(2013)}]{alexakis13}%
  \BibitemOpen
  \bibfield  {author} {\bibinfo {author} {\bibfnamefont {A.}~\bibnamefont
  {Alexakis}},\ }\href {\doibase 10.1103/PhysRevLett.110.084502} {\bibfield
  {journal} {\bibinfo  {journal} {Phys. Rev. Lett.}\ }\textbf {\bibinfo
  {volume} {110}},\ \bibinfo {pages} {084502} (\bibinfo {year}
  {2013})}\BibitemShut {NoStop}%
\bibitem [{\citenamefont {Nazarenko}(2011)}]{nazarenko11}%
  \BibitemOpen
  \bibfield  {author} {\bibinfo {author} {\bibfnamefont {S.}~\bibnamefont
  {Nazarenko}},\ }\href@noop {} {\emph {\bibinfo {title} {Wave turbulence}}},\
  Vol.\ \bibinfo {volume} {825}\ (\bibinfo  {publisher} {Springer Verlag},\
  \bibinfo {year} {2011})\BibitemShut {NoStop}%
\bibitem [{\citenamefont {Iroshnikov}(1964)}]{iroshnikov64}%
  \BibitemOpen
  \bibfield  {author} {\bibinfo {author} {\bibfnamefont {P.~S.}\ \bibnamefont
  {Iroshnikov}},\ }\href@noop {} {\bibfield  {journal} {\bibinfo  {journal}
  {Soviet Astronomy}\ }\textbf {\bibinfo {volume} {7}},\ \bibinfo {pages} {566}
  (\bibinfo {year} {1964})}\BibitemShut {NoStop}%
\bibitem [{\citenamefont {Kraichnan}(1965)}]{kraichnan65}%
  \BibitemOpen
  \bibfield  {author} {\bibinfo {author} {\bibfnamefont {R.~H.}\ \bibnamefont
  {Kraichnan}},\ }\href {\doibase 10.1063/1.1761412} {\bibfield  {journal}
  {\bibinfo  {journal} {Phys. Fluids}\ }\textbf {\bibinfo {volume} {8}},\
  \bibinfo {pages} {1385} (\bibinfo {year} {1965})}\BibitemShut {NoStop}%
\bibitem [{\citenamefont {Taylor}(1935)}]{taylor35}%
  \BibitemOpen
  \bibfield  {author} {\bibinfo {author} {\bibfnamefont {G.~I.}\ \bibnamefont
  {Taylor}},\ }\href@noop {} {\bibfield  {journal} {\bibinfo  {journal}
  {Proceedings of the Royal Society of London. Series A, Mathematical and
  Physical Sciences}\ }\textbf {\bibinfo {volume} {151}},\ \bibinfo {pages}
  {421} (\bibinfo {year} {1935})}\BibitemShut {NoStop}%
\bibitem [{\citenamefont {Bec}\ and\ \citenamefont
  {Khanin}(2007)}]{beckhanin07}%
  \BibitemOpen
  \bibfield  {author} {\bibinfo {author} {\bibfnamefont {J.}~\bibnamefont
  {Bec}}\ and\ \bibinfo {author} {\bibfnamefont {K.}~\bibnamefont {Khanin}},\
  }\href {\doibase 10.1016/j.physrep.2007.04.002} {\bibfield  {journal}
  {\bibinfo  {journal} {Phys. Reports}\ }\textbf {\bibinfo {volume} {447}},\
  \bibinfo {pages} {1} (\bibinfo {year} {2007})}\BibitemShut {NoStop}%
\bibitem [{\citenamefont {Parker}(1994)}]{parker94}%
  \BibitemOpen
  \bibfield  {author} {\bibinfo {author} {\bibfnamefont {E.~N.}\ \bibnamefont
  {Parker}},\ }\href@noop {} {\emph {\bibinfo {title} {Spontaneous current
  sheets in magnetic fields: with applications to stellar x-rays}}}\ (\bibinfo
  {publisher} {Oxford University Press},\ \bibinfo {year} {1994})\BibitemShut
  {NoStop}%
\bibitem [{\citenamefont {Greco}\ \emph {et~al.}(2010)\citenamefont {Greco},
  \citenamefont {Servidio}, \citenamefont {Matthaeus},\ and\ \citenamefont
  {Dmitruk}}]{grecoetal10}%
  \BibitemOpen
  \bibfield  {author} {\bibinfo {author} {\bibfnamefont {A.}~\bibnamefont
  {Greco}}, \bibinfo {author} {\bibfnamefont {S.}~\bibnamefont {Servidio}},
  \bibinfo {author} {\bibfnamefont {W.}~\bibnamefont {Matthaeus}}, \ and\
  \bibinfo {author} {\bibfnamefont {P.}~\bibnamefont {Dmitruk}},\ }\href
  {\doibase 10.1016/j.pss.2010.08.019} {\bibfield  {journal} {\bibinfo
  {journal} {Planetary and Space Science}\ }\textbf {\bibinfo {volume} {58}},\
  \bibinfo {pages} {1895} (\bibinfo {year} {2010})}\BibitemShut {NoStop}%
\bibitem [{\citenamefont {Li}(2008)}]{li08}%
  \BibitemOpen
  \bibfield  {author} {\bibinfo {author} {\bibfnamefont {G.}~\bibnamefont
  {Li}},\ }\href@noop {} {\bibfield  {journal} {\bibinfo  {journal} {The
  Astrophysical Journal Letters}\ }\textbf {\bibinfo {volume} {672}},\ \bibinfo
  {pages} {L65} (\bibinfo {year} {2008})}\BibitemShut {NoStop}%
\bibitem [{\citenamefont {Tsinober}(2009)}]{tsinober09}%
  \BibitemOpen
  \bibfield  {author} {\bibinfo {author} {\bibfnamefont {A.}~\bibnamefont
  {Tsinober}},\ }\href@noop {} {\emph {\bibinfo {title} {An informal conceptual
  introduction to turbulence}}}\ (\bibinfo  {publisher} {Springer},\ \bibinfo
  {year} {2009})\BibitemShut {NoStop}%
\bibitem [{\citenamefont {Jimenez}\ \emph {et~al.}(1993)\citenamefont
  {Jimenez}, \citenamefont {Wray}, \citenamefont {Saffman},\ and\ \citenamefont
  {Rogallo}}]{jimenezetal93}%
  \BibitemOpen
  \bibfield  {author} {\bibinfo {author} {\bibfnamefont {J.}~\bibnamefont
  {Jimenez}}, \bibinfo {author} {\bibfnamefont {A.~A.}\ \bibnamefont {Wray}},
  \bibinfo {author} {\bibfnamefont {P.~G.}\ \bibnamefont {Saffman}}, \ and\
  \bibinfo {author} {\bibfnamefont {R.~S.}\ \bibnamefont {Rogallo}},\ }\href
  {\doibase 10.1017/S0022112093002393} {\bibfield  {journal} {\bibinfo
  {journal} {J. Fluid Mech.}\ }\textbf {\bibinfo {volume} {255}},\ \bibinfo
  {pages} {65} (\bibinfo {year} {1993})}\BibitemShut {NoStop}%
\bibitem [{\citenamefont {Holmes}\ \emph {et~al.}(1998)\citenamefont {Holmes},
  \citenamefont {Lumley},\ and\ \citenamefont {Berkooz}}]{holmesetal98}%
  \BibitemOpen
  \bibfield  {author} {\bibinfo {author} {\bibfnamefont {P.}~\bibnamefont
  {Holmes}}, \bibinfo {author} {\bibfnamefont {J.~L.}\ \bibnamefont {Lumley}},
  \ and\ \bibinfo {author} {\bibfnamefont {G.}~\bibnamefont {Berkooz}},\
  }\href@noop {} {\emph {\bibinfo {title} {{Turbulence, coherent structures,
  dynamical systems and symmetry}}}}\ (\bibinfo  {publisher} {Cambridge
  University Press},\ \bibinfo {year} {1998})\BibitemShut {NoStop}%
\end{thebibliography}%

\end{document}